\documentclass[twocolumn,tighten,twocolappendix]{aastex63}

\usepackage{amssymb,amsmath,amsthm}
\usepackage{graphicx}
\usepackage{mathrsfs}
\usepackage{booktabs}
\usepackage{multirow}
\usepackage{empheq}
\usepackage{verbatim}
\usepackage{mathtools}
\usepackage{hyperref}
\usepackage{xcolor}
\usepackage{xspace}
\usepackage{lineno}
\usepackage[makeroom]{cancel}
\usepackage[acronym]{glossaries}

\def\kms{km\,s\ensuremath{^{-1}}\xspace} % km/s

\begin{document}

\title{183\,GHz water megamasers in active galactic nuclei: a new accretion disk tracer}

\correspondingauthor{Dominic~W.~Pesce}
\email{dpesce@cfa.harvard.edu}

\author[0000-0002-5278-9221]{Dominic~W.~Pesce}
\affiliation{Center for Astrophysics $|$ Harvard \& Smithsonian, 60 Garden Street, Cambridge, MA 02138, USA}
\affiliation{Black Hole Initiative at Harvard University, 20 Garden Street, Cambridge, MA 02138, USA}

\author[0000-0002-1468-9203]{James~A.~Braatz}
\affiliation{National Radio Astronomy Observatory, 520 Edgemont Road, Charlottesville, VA 22903, USA}
%\email{jbraatz@nrao.edu}

\author[0000-0002-7495-4005]{Christian Henkel}
\affiliation{Max-Planck-Institut f\"ur Radioastronomie, Auf dem H\"ugel 69, D-53121 Bonn, Germany}
\affiliation{Astronomy Department, Faculty of Science, King Abdulaziz University, P.O. Box 80203, Jeddah 21589, Saudi Arabia}
%\email{chenkel@mpifr-bonn.mpg.de}

\author[0000-0001-9549-6421]{Elizabeth M. L. Humphreys}
\affiliation{Joint ALMA Observatory, Alonso de Cordova 3107, Vitacura, Santiago, Chile}
\affiliation{European Southern Observatory (ESO) Vitacura, Alonso de Cordova 3107, Vitacura, Santiago, Chile}

\author[0000-0002-3443-2472]{C. M. Violette Impellizzeri}
\affiliation{Leiden Observatory, Leiden University, PO Box 9513, 2300 RA Leiden, The Netherlands }
\affiliation{National Radio Astronomy Observatory, 520 Edgemont Road, Charlottesville, VA 22903, USA}

\author[0000-0001-6211-5581]{Cheng-Yu Kuo}
\affiliation{Institute of Astronomy and Astrophysics, Academia Sinica, 11F of Astronomy-Mathematics Building, AS/NTU No. 1, Sec. 4, Roosevelt Rd, Taipei 10617, Taiwan, R.O.C.}
\affiliation{Physics Department, National Sun Yat-Sen University, No. 70, Lien-Hai Road, Kaosiung City 80424, Taiwan, R.O.C.}

\begin{abstract}
We present the results of an ALMA survey to identify 183\,GHz H$_2$O maser emission from AGN already known to host 22\,GHz megamaser systems.  Out of 20 sources observed, we detect significant 183\,GHz maser emission from 13; this survey thus increases the number of AGN known to host (sub)millimeter megamasers by a factor of 5.  We find that the 183\,GHz emission is systematically fainter than the 22\,GHz emission from the same targets, with typical flux densities being roughly an order of magnitude lower at 183\,GHz than at 22\,GHz.  However, the isotropic luminosities of the detected 183\,GHz sources are comparable to their 22\,GHz values.  For two of our sources -- ESO 269-G012 and the Circinus galaxy -- we detect rich 183\,GHz spectral structure containing multiple line complexes.  The 183\,GHz spectrum of ESO 269-G012 exhibits the triple-peaked structure characteristic of an edge-on AGN disk system.  The Circinus galaxy contains the strongest 183\,GHz emission detected in our sample, peaking at a flux density of nearly 5\,Jy.  The high signal-to-noise ratios achieved by these strong lines enable a coarse mapping of the 183\,GHz maser system, in which the masers appear to be distributed similarly to those seen in VLBI maps of the 22\,GHz system in the same galaxy and may be tracing the circumnuclear accretion disk at larger orbital radii than are occupied by the 22\,GHz masers.  This newly identified population of AGN disk megamasers presents a motivation for developing VLBI capabilities at 183\,GHz.
\end{abstract}

\section{Introduction}

H$_2$O megamasers residing in the accretion disks of active galactic nuclei (AGN) have proven to be unique tools for mapping circumnuclear molecular gas on sub-pc scales \citep[e.g., ][]{Greenhill_1995,Argon_2007,Reid_2009}, providing precise constraints on supermassive black hole (SMBH) masses \citep[e.g., ][]{Kuo_2011,Gao_2017,Zhao_2018}, and enabling geometric distance measurements to their host galaxies \citep[e.g.,][]{Herrnstein_1999,Braatz_2010,Humphreys_2013,Kuo_2013,Gao_2016,Reid_2019,Pesce_2020}.  For galaxies situated in the Hubble flow, these distance measurements can be used to provide direct constraints on the local cosmological expansion rate \citep{Reid_2013,Kuo_2015,Pesce_2020b}.

To date, most of the observational work on AGN disk megamasers has focused on the $6_{1,6}$-$5_{2,3}$ rotational transition in the ground vibrational state of the ortho-H$_2$O molecule (i.e., the version of the molecule in which the nuclear spins of the two hydrogen atoms are parallel).  This transition emits at a rest-frame frequency of $\sim$22.23508\,GHz\footnote{In astrophysical systems, the observed $6_{1,6}$-$5_{2,3}$ line is typically composed of a blend of six hyperfine transitions spanning $\sim$0.5\,MHz; the 22.23508\,GHz nominal rest-frame frequency represents an intensity-weighted (assuming thermal equilibrium) mean of these six transitions.} \citep{Kukolich_1969}, though we adopt the community standard and refer to it in this paper as simply ``the 22\,GHz line.''  Nearly 200 galaxies have been detected in this line so far -- the result of more than 4800 galaxies surveyed \citep{Kuo_2018,Kuo_2020} -- and $\sim$150 of these detections are associated with AGN.\footnote{\url{https://safe.nrao.edu/wiki/bin/view/Main/PublicWaterMaserList}}

A $\sim$20\% subset of these AGN megamasers originate from circumnuclear accretion disks and are identifiable by their characteristic spectral profiles containing three distinct groups of maser features distributed approximately symmetrically around the recession velocity of the host galaxy \citep[see, e.g.,][]{Pesce_2015}.  The central group of features (called the ``systemic masers'') coincides roughly with the recession velocity of the galaxy; these features originate from masing gas that resides along our line of sight through the disk to the central AGN.  The two other groups of features -- which are typically offset by several hundred \kms redward and blueward of the galaxy recession velocity -- are collectively known as the ``high-velocity'' features and originate from masing gas on lines of sight through the disk that are tangent to its orbital motion.

Though the 22\,GHz megamaser transition is by far the best studied, the H$_2$O molecule has a rich energy spectrum containing many other transitions capable of sustaining maser activity under physical conditions comparable to those that support 22\,GHz emission \citep{Neufeld_1991,Yates_1997,Gray_2016}.  To date, there have been only three AGN observed to host H$_2$O megamaser emission at frequencies other than 22\,GHz:
\begin{itemize}
    \item The first detection of an extragalactic H$_2$O maser source in a transition other than the 22\,GHz line was made by \cite{Humphreys_2005}, who used the Submillimeter Array (SMA) to observe 183\,GHz megamaser emission from NGC 3079.  The 183\,GHz emission from NGC 3079 is consistent with an origin near the AGN and has a flux density about $\sim$3 times fainter than the corresponding 22\,GHz emission; a tentative detection of 439\,GHz maser emission with the James Clerk Maxwell Telescope (JCMT) towards the same galaxy was also identified by the authors.
    \item \cite{Hagiwara_2013} used the Atacama Large Millimeter/submillimeter Array (ALMA) to detect 321\,GHz H$_2$O megamaser emission towards the Circinus galaxy, where it is observed to be roughly two orders of magnitude fainter than the corresponding 22\,GHz emission.  The spectral structure at 321\,GHz is similar to that at 22\,GHz, and repeated observations have demonstrated that it persists over years but exhibits large ($\sim$order of magnitude) variability in the maser line strength \citep{Pesce_2016,Hagiwara_2021}.
    \item \cite{Humphreys_2016} used the Atacama Pathfinder Experiment (APEX) telescope to detect 183\,GHz megamaser emission from NGC 4945, where it is observed to have a flux density that is a factor of $\sim$1.5 times greater than that of the corresponding 22\,GHz emission.  \cite{Pesce_2016} and \cite{Hagiwara_2016} additionally identified 321\,GHz megamaser emission from NGC 4945 using ALMA observations, at a level more than two orders of magnitude fainter than the 22\,GHz or 183\,GHz emission but with a similar spectral structure.  The 321\,GHz maser system appears to be time-variable, as followup ALMA observations failed to detect any emission from NGC 4945 \citep{Hagiwara_2021}.
\end{itemize}
\noindent The 183\,GHz transition has also been detected towards $\sim$10 ultraluminous infrared galaxies (ULIRGs), where it is typically associated with spatially distributed star formation rather than centrally concentrated AGN activity \citep{Cernicharo_2006,Konig_2017,Imanishi_2022}.  One possible exception is the so-called ``Superantennae'' pair of interacting galaxies, which are ULIRGs but for which the 183\,GHz emission appears to be at least partially circumnuclear and has been associated with an AGN \citep{Imanishi_2021}.

In this paper we present the results of an ALMA survey that increases by a factor of 5 the number of AGN known to host H$_2$O megamaser emission at a frequency other than 22\,GHz.  Our survey targets the $3_{1,3}$-$2_{2,0}$ rotational transition in the ground vibrational state of the para-H$_2$O molecule (where the nuclear spins of the two hydrogen atoms are anti-parallel), which has a rest-frame emission frequency of 183.310087\,GHz\footnote{\url{https://splatalogue.online/}} \citep{Pickett_1998}.  We refer to this transition as ``the 183\,GHz line.''

This paper is organized as follows.  In \autoref{sec:Reduction} we describe the observations, data reduction, and imaging procedure.  \autoref{sec:Results} presents the results of our survey and compares the 183\,GHz spectra to prior 22\,GHz observations.  We summarize and conclude in \autoref{sec:Summary}.  Throughout this paper, all velocities are quoted using the optical convention in the heliocentric reference frame unless otherwise specified, and we adopt a Hubble constant value of $H_0 = 70$\,km\,s$^{-1}$\,Mpc$^{-1}$.

\section{Data reduction and imaging} \label{sec:Reduction}

The data presented in this paper were taken as part of ALMA projects 2017.1.00909.S and 2018.1.00321.S.  We used ALMA's Band 5 to observe 20 AGN known to host 22\,GHz water megamaser systems and spanning a range of redshifts out to $z \approx 0.05$; the targets are listed in \autoref{tab:Observations}.  Each of the targeted galaxies was selected either because it was identified as a disk maser by \citet{Pesce_2015}, and/or -- as in the case of NGC 1068, NGC 1386, Circinus, and NGC 5643 -- because it is a strong 22\,GHz megamaser source associated with an AGN.  Observations of the science targets were interwoven with observations of calibrator targets, which are used to determine bandpass, flux density, and station gain calibrations.  All observations recorded only dual linear polarization correlation products (XX, YY), and cross-hand data are not available.

The spectral setup for each observation consisted of four spectral windows.  Three of the spectral windows were configured to have fine frequency resolution (122\,kHz, corresponding to $\sim$0.2\,\kms) and to span a total of $\pm$ $\sim$1200\,\kms around the expected locations of the maser lines (i.e., centered on the recession velocity of each galaxy) with a small ($\sim$50\,\kms) overlap to ensure continuous coverage.  The fourth spectral window was wider (2\,GHz, corresponding to $\sim$3200\,\kms) with coarser channels (15.625\,MHz, corresponding to $\sim$25\,\kms), and was offset toward lower frequency to capture continuum emission.

Though most of the observations taken as part of these ALMA projects have corresponding quality assurance 2 (QA2) data processing, we also include a handful of observations that did not achieve sufficient sensitivity to pass the QA2 level.  Additionally, there are some desirable data reduction procedures (e.g., self-calibration) that are not included in the standard QA2 pipeline.  To maintain uniformity, we thus carry out our own data reduction on all datasets.  Our data reduction procedure for each observation largely follows standard practice for ALMA datasets.  We have carried out all data reduction for this paper using CASA\footnote{\url{https://casa.nrao.edu/}} version 6.2.1.7.  Where the calibration procedure for individual science targets deviates from the general procedure described in this section, or where there is relevant information specific to an individual source, we have provided additional details in \autoref{app:Calibration}.

\begin{deluxetable*}{lcccc}
\tablecolumns{5}
\tablewidth{0pt}
\tablecaption{Observation details\label{tab:Observations}}
\tablehead{\colhead{Science target} & \colhead{Right ascension} & \colhead{Declination} & \colhead{$v_{\text{rec}}$ (\kms)} & \colhead{Observing date(s)}}
\startdata
J0109-0332 & 01:09:45.101 & $-03$:32:32.97 & 16369 & 2018 Nov 06 \\
J0126-0417 & 01:26:01.640 & $-04$:17:56.42 & 5639 & 2018 Nov 16 \\
NGC 1068 & 02:42:40.771 & $-00$:00:47.84 & 1137 & 2018 Nov 9, Nov 16 \\
NGC 1194 & 03:03:49.100 & $-01$:06:12.99 & 4086 & 2018 Nov 16 \\
NGC 1386 & 03:36:46.238 & $-35$:59:57.39 & 868 & 2018 Nov 16 \\
ESO 558-G009 & 07:04:21.002 & $-21$:35:19.36 & 7674 & 2018 Oct 31 \\
IC 485 & 08:00:19.772 & $+26$:42:05.35 & 8342 & 2018 Oct 31, Dec 06 \\
J0847-0022 & 08:47:47.692 & $-00$:22:51.47 & 15243 & 2018 Nov 08 \\
Mrk 1419 & 09:40:36.376 & $+03$:34:36.96 & 4947 & 2018 Nov 14 \\
IC 2560 & 10:16:18.666 & $-33$:33:49.85 & 2925 & 2018 Oct 02, Oct 30, Nov 08 \\
NGC 3393 & 10:48:23.467 & $-25$:09.43.30 & 3750 & 2018 Oct 30, Nov 14 \\
UGC 6093 & 11:00:47.955 & $+10$.43.41.76 & 10803 & 2018 Oct 31 \\
ESO 269-G012 & 12:56:40.521 & $-46$:55:33.79 & 5014 & 2018 Dec 01 \\
CGCG 074-064 & 14:03:04.459 & $+08$:56:51.03 & 6915 & 2018 Dec 06 \\
NGC 5495 & 14:12:23.353 & $-27$:06:28.72 & 6737 & 2018 Dec 06 \\
Circinus & 14:13:09.950 & $-65$:20:21.20 & 434 & 2018 Dec 01 \\
NGC 5643 & 14:32:40.778 & $-44$:10:28.60 & 1199 & 2018 Aug 27, Sep 20 \\
NGC 5765b & 14:50:51.500 & $+05$:06:52.00 & 8257 & 2018 Dec 06 \\
CGCG 165-035 & 15:14:39.793 & $+26$:35:39.11 & 9622 & 2018 Dec 06 \\
NGC 6264 & 16:57:16.110 & $+27$:50:58.71 & 10145 & 2018 Nov 18 \\
\enddata
\tablecomments{Information about the targets surveyed in this paper for 183\,GHz emission.  Coordinates are specified in J2000 and indicate the pointing center for the science targets; observing dates are specified in UT.  Recession velocities $v_{\text{rec}}$ have been obtained from the NASA/IPAC Extragalactic Database (NED).}
\end{deluxetable*}

\subsection{Data reduction procedure} \label{sec:Calibration}

We start by importing the raw ALMA Science Data Model files and converting them to CASA Measurement Sets (MSs) using the \texttt{importasdm} task.  Initial flagging is carried out using the \texttt{flagdata} task; we remove pointing scans, autocorrelation products, and any data associated with shadowed antennas.  We use the \texttt{gencal} task to create system temperature and water vapor radiometer calibration tables, which we then apply to the science and calibrator targets using the \texttt{applycal} task.  We then perform a round of data inspection and manual flagging, the latter typically restricted to edge channels of the continuum spectral windows and to the initial integrations of some scans.

Prior to determining the bandpass solution, we first track the phase variations of the bandpass calibrator on every integration time by averaging over a small number of frequency channels ($\sim$5\% of each spectral window) using the \texttt{gaincal} task.  We select as our reference antenna the one that is most centrally located within the array configuration while also being present for all scans, and phase solutions are referenced to the phase at this antenna.  These phase solutions are then applied to the bandpass calibrator prior to deriving the phase and amplitude bandpass solutions on a per-channel basis using the \texttt{bandpass} task.

We determine the absolute flux density scale using observations of targets from the ALMA calibrator source catalog\footnote{\url{https://almascience.nrao.edu/sc/}}.  For each flux density calibrator, we assume a power-law spectral energy distribution to extrapolate from the ALMA calibrator source catalog measurements to the $\sim$183\,GHz observing frequencies relevant for the observations in this paper.  Specifically, we use
\begin{equation}
S_{\nu} = S_{\nu,0} \left( \frac{\nu}{\nu_0} \right)^{\alpha} ,
\end{equation}
\noindent where $S_{\nu}$ is the flux density at frequency $\nu$, $S_{\nu,0}$ is the flux density at a reference frequency $\nu_0$, and $\alpha$ is the spectral index.  We specify the absolute flux density calibration using CASA's \texttt{setjy} task.  The flux densities and spectral indices assumed for each of our calibrators are detailed in \autoref{app:Calibration}.

We carry out gain calibration in a few steps.  Applying bandpass calibration on-the-fly, we first use the \texttt{gaincal} task to derive gain phase solutions on the calibrator targets per integration time.  We separately solve for the gain phases on the calibrator targets per scan time, for later application to the science target.  We apply the integration-time gain phase solutions to the calibrator targets during another run of \texttt{gaincal} in which we solve for both the gain amplitudes and phases per scan time.  Following this gain calibration, we bootstrap the absolute flux density calibration from the flux density calibrator to the gain calibrator using the \texttt{fluxscale} task.  The gain phase and amplitude solutions -- the latter of which now carry over the absolute flux density calibration -- are then applied to the science target using the \texttt{applycal} task.

\subsection{Imaging procedure} \label{sec:Imaging}

We carry out all imaging for this paper using the H\"ogbom CLEAN method \citep{Hogbom_1974} as implemented in CASA's \texttt{tclean} task, combining the calibrated XX and YY visibilities to form Stokes I images.  We use the default ``standard'' $(u,v)$-gridding setting in \texttt{tclean}, and we specify natural visibility weighting to maximize the point source sensitivity of our images.  Each image is reconstructed on a $300 \times 300$ pixel field of view, with pixel sizes chosen to be equal to ${\sim}1/8$ of the nominal beam FWHM.  We do not place any CLEAN boxes, and instead we permit the algorithm to operate on the entire field of view.  We use a loop gain of 0.1, and as our stopping criterion we require that the peak residual be $\sim$5 times the expected image RMS level.

We carry out imaging separately for every channel in each spectral window to produce image cubes, and we also carry out continuum imaging by combining multiple channels to form a single image.  For sources with detected 183\,GHz spectral line emission (see \autoref{sec:Results}), we carry out continuum imaging using only the continuum spectral window.  For sources with no detected spectral lines, we combine all spectral windows during continuum imaging.  The spectra and continuum images for each of our targets are shown in \autoref{app:Images}.  For the two sources with other molecular line emission, we instead perform continuum imaging using only those channels in the continuum spectral window that lack spectral line emission; the images of these other molecular lines are shown in \autoref{app:HCNetc}.

\subsection{Spectral line identification} \label{sec:Identification}

We manually inspect spectra extracted from the central region of each of the image cubes for signs of spectral line emission, both before and after carrying out boxcar smoothing in frequency by 10 channels (corresponding to a $\sim$2\,\kms post-averaging spectral resolution).  For some targets, the spectral line emission is sufficiently strong for the detection to be unambiguous.  Other targets have weaker emission, and so we establish some detection criteria below.

After initially estimating the spectral sensitivity by computing the RMS across the full spectral window\footnote{We note that for our lowest-redshift targets (e.g., Circinus, NGC 1386), the steep transmission gradient caused by the atmospheric 183\,GHz line causes the the RMS noise level to change substantially across the spectral windows.  We do not attempt to compensate for this effect when computing the RMS, so our estimate is an average value across the entire spectral range.}, we identify any peaks that reach the ${\gtrsim}5\sigma$ level.  If the source shows comparably elevated emission across multiple channels, then we consider it to be a detection.

\section{Results and discussion} \label{sec:Results}

In total, we detect 183\,GHz spectral line emission in 13 out of 20 targeted sources; \autoref{tab:Detections} provides a summary of the targets detected in 183\,GHz line emission and in continuum, and at what sensitivities.  The detected 183\,GHz spectra are shown in Figures \ref{fig:hex1}, \ref{fig:hex2}, and \ref{fig:Circinus}, where they are also compared against 22\,GHz spectra.  The full set of 183\,GHz spectra and continuum images for each of our targets is shown in \autoref{app:Images}; for sources detected in both spectral line and continuum emission, we typically find that the line and continuum emission are spatially coincident.

Because maser sources are expected to be heavily anisotropic emitters, the intrinsic luminosity of a maser is difficult to estimate without knowledge of the solid angle into which the emission is beamed.  Instead, we characterize the strength of a maser system using the ``isotropic luminosity,'' $L_{\text{iso}}$, which corresponds to the luminosity that an isotropic emitter would need to have to produced the observed flux level.  We compute isotropic luminosities using

\begin{equation}
L_{\text{iso}} = \frac{4 \pi D^2 \nu_0}{c} \int S_{v} dv ,
\end{equation}

\noindent where $D$ is the distance to the system, $\nu_0$ is the rest-frame frequency of the emission line, and $S_{v}$ is the flux density as a function of velocity $v$.  Scaled to convenient units for 183\,GHz maser emission, this equation becomes

\begin{equation}
L_{\text{iso}} = \left( 1.9 \times 10^{-4} \text{ L}_{\odot} \right) \left( \frac{D}{1 \text{ Mpc}} \right)^2 \left( \frac{\int S_{v} dv}{1 \text{ mJy \kms}} \right) .
\end{equation}

\noindent The true intrinsic luminosity is smaller than $L_{\text{iso}}$ by a factor $\Omega / 4\pi$, where $\Omega$ is the solid angle of the maser beam.

In this section, we describe the properties of each detection, and we compare the 183\,GHz maser line characteristics to those of the 22\,GHz masers in the same system.  We note that because these megamaser sources are known to be intrinsically variable on timescales of $\sim$weeks at 22\,GHz \citep{Lo_2005}, the typically large temporal separation between the observations of the 183\,GHz masers presented in this paper and their 22\,GHz counterparts from the literature renders a line-by-line comparison of the two spectra potentially fraught.  Instead, we opt here for a somewhat more qualitative comparison, aiming to determine whether the two sets of spectra exhibit comparable gross structures.

\begin{deluxetable*}{lccccc}
\tablecolumns{6}
\tablewidth{0pt}
\tablecaption{Detection details\label{tab:Detections}}
\tablehead{ & \colhead{Spectral sensitivity} & \colhead{Continuum sensitivity}  & \colhead{Peak 183\,GHz} & \colhead{Peak continuum} & \colhead{Isotropic luminosity$^*$} \\
\colhead{Science target} & \colhead{(RMS mJy)} & \colhead{(RMS mJy\,arcsec$^{-2}$)} & \colhead{(mJy)} & \colhead{(mJy\,arcsec$^{-2}$)} & \colhead{(L$_{\odot}$)}}
\startdata
J0109-0332 & 1.6 & 0.26 & no detection & 4.80 & \ldots \\
J0126-0417 & 1.9 & 0.28 & no detection & 1.17 & \ldots \\
NGC 1068 & 7.4 & 1.13 & no detection & 128 & \ldots \\
NGC 1194 & 2.2 & 0.34 & 35.2$^\dag$ & 11.1 & $64 \pm 12$ \\
NGC 1386 & 9.3 & 0.67 & no detection & 22.7 & \ldots \\
ESO 558-G009 & 2.2 & 0.35 & 11.8 & 1.67 & $402 \pm 41$ \\
IC 485 & 1.6 & 0.15 & 8.4 & 1.15 & $527 \pm 57$ \\
J0847-0022 & 1.2 & 0.24 & 6.3 & 1.46 & $1180 \pm 112$ \\
Mrk 1419 & 2.2 & 0.29 & 23.6 & 1.07 & $542 \pm 71$ \\
IC 2560 & 2.8 & 0.25 & 29.6 & 5.42 & $282 \pm 59$ \\
NGC 3393 & 1.8 & 0.21 & no detection & 4.23 & \ldots \\
UGC 6093 & 1.9 & 0.32 & 16.0 & no detection & $978 \pm 86$ \\
ESO 269-G012 & 2.0 & 0.09 & 30.8 & no detection & $1337 \pm 165$ \\
CGCG 074-064 & 1.9 & 0.08 & 17.9 & 0.28 & $484 \pm 88$ \\
NGC 5495 & 2.1 & 0.11 & 9.9 & no detection & $163 \pm 27$ \\
Circinus$^\ddag$ & 21.8 & 0.42 & 4810 & 90.2 & $165 \pm 63$ \\
NGC 5643 & 4.9 & 1.08 & 32.9 & 14.2 & $27 \pm 3$ \\
NGC 5765b & 1.3 & 0.05 & 6.5 & 0.73 & $379 \pm 75$ \\
CGCG 165-035 & 1.9 & 0.33 & no detection & no detection & \ldots \\
NGC 6264 & 2.0 & 0.23 & no detection & 0.78 & \ldots \\
\enddata
\tablecomments{The RMS spectral sensitivity and peak 183\,GHz line flux density is determined after boxcar smoothing by 10 channels to achieve a $\sim$2\,\kms velocity resolution.  Spectral line detections are determined by the criteria described in \autoref{sec:Identification}; continuum nondetections are only reported if the emission level at the expected location of the AGN does not rise above the $3\sigma$ level.\\
$^*$When computing the uncertainty in isotropic luminosity, we incorporate an assumed 300\,\kms peculiar-velocity-induced uncertainty in the recession velocity for galaxies whose distances are estimated using the Hubble law.\\
$^\dag$The peak flux density of the NGC 1194 spectrum occurs in a single narrow (FWHM $<$1\,\kms) line that is offset from the systemic velocity by $\sim$500\,\kms (see \autoref{sec:NGC1194}), so we report the flux density here prior to smoothing.  The peak flux density after smoothing over 10 channels occurs in the systemic set of features and reaches a value of $\sim$16\,mJy.\\
$^\ddag$Because the Circinus spectrum contains many narrow ($\lesssim$1\,\kms) lines, we report the RMS spectral sensitivity and peak 183\,GHz line flux density at the native channel resolution of $\sim$0.2\,\kms (i.e., with no smoothing applied).}
\end{deluxetable*}

\begin{figure*}
    \centering
    \includegraphics[width=1.00\textwidth]{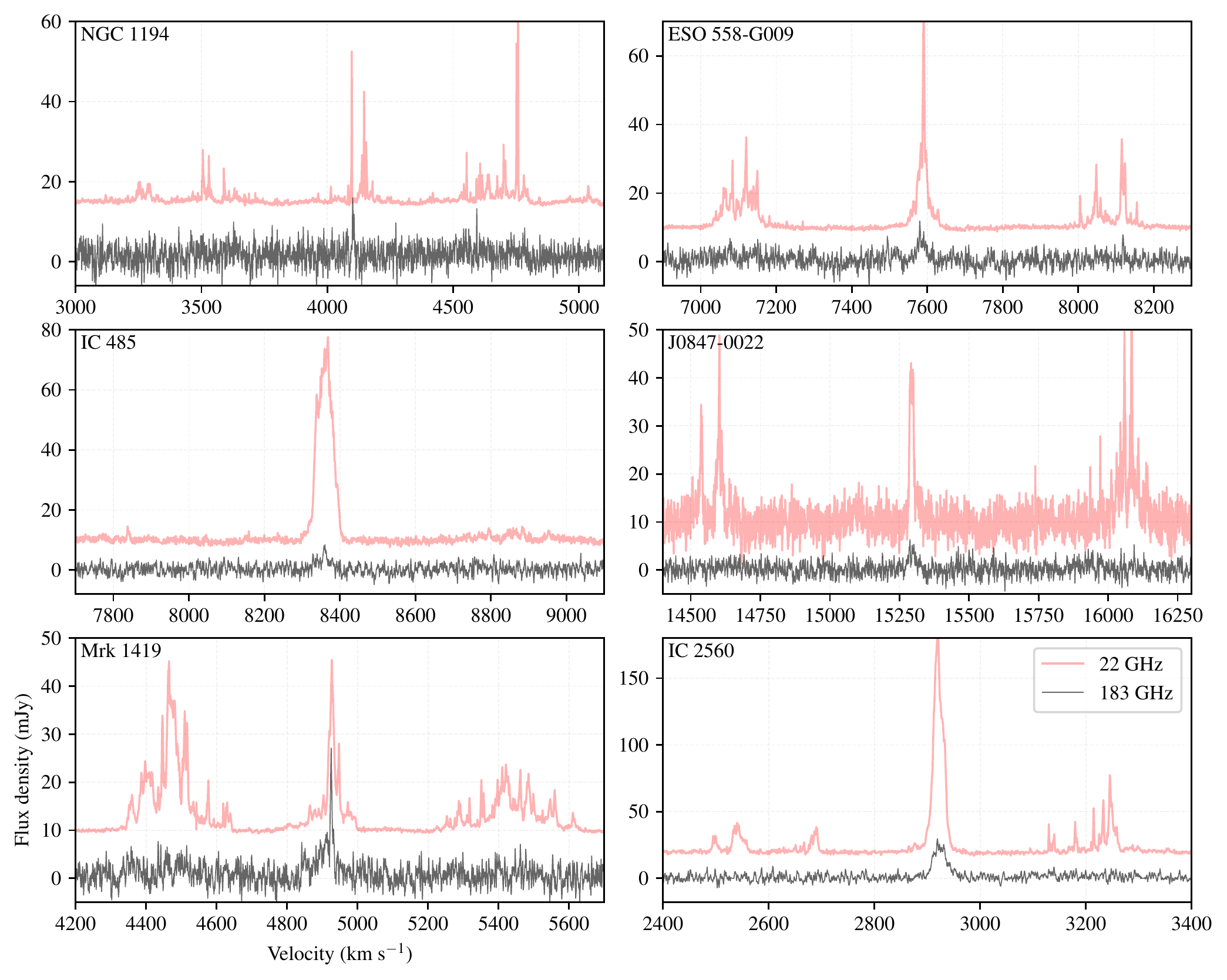}
    \caption{Comparison of 183\,GHz spectra (in black) with their 22\,GHz counterparts (in red, vertically offset).  The 183\,GHz spectra have been boxcar smoothed across 10 channels, corresponding to a $\sim$2\,\kms post-averaging spectral resolution.  Note that the pair of spectra in each panel were not observed concurrently; in most cases, the 22\,GHz spectrum predates the 183\,GHz spectrum by multiple years.}\label{fig:hex1}
\end{figure*}

\subsection{NGC 1194} \label{sec:NGC1194}

The 22\,GHz maser system in NGC 1194 was discovered by the Megamaser Cosmology Project (MCP), and VLBI observations presented in \citet{Kuo_2011} confirm that it is a disk maser system.  Relative to other 22\,GHz disk maser systems, NGC 1194 is notable for its large apparent disk size ($\sim$0.54-1.33\,pc across; see \citealt{Kuo_2011}) and for having several strong (up to $\sim$1\,Jy) and narrow (linewidths $<$1\,\kms) individual maser lines exhibiting a large degree of variability \citep{Pesce_2015}.  Our 183\,GHz spectrum of NGC 1194 is shown in the upper lefthand panel of \autoref{fig:hex1}, along with a 22\,GHz spectrum reproduced from \citet{Pesce_2015} (see Fig. 1 in that paper).  The 22\,GHz spectrum is an average of multiple epochs of spectral monitoring observations taken with the Green Bank Telescope (GBT).

The most prominent component in the 183\,GHz spectrum of NGC 1194 is the central set of systemic maser features, which span a velocity range of $\sim$10\,\kms around a central velocity of ${\sim}4104$\,\kms.  The peak flux density of the systemic features is $\sim$16\,mJy, which is a factor of $\sim$2 fainter than the typical peak flux density of the 22\,GHz systemic masers.

A second component in the 183\,GHz spectrum of NGC 1194 is a single narrow ($\text{FWHM} \approx 0.69$\,\kms) line at a velocity of $\sim$4593.5\,\kms, which peaks at a flux density of $\sim$35.2\,mJy.  This spectral line coincides with the redshifted complex of high-velocity maser features seen at 22\,GHz.

Adopting a Hubble law distance of 58\,Mpc, the isotropic luminosity of the 183\,GHz maser emission from NGC 1194 is $L_{\text{iso}} \approx 64$\,L$_{\odot}$.  This value is a factor of $\sim$2 lower than the $\sim$130\,L$_{\odot}$ isotropic luminosity observed at 22\,GHz \citep{Pesce_2015}.

\subsection{ESO 558-G009}

The 22\,GHz maser system in ESO 558-G009 was discovered by the MCP, and VLBI observations presented in \citet{Gao_2017} confirm that it is a disk maser system.  Our 183\,GHz spectrum of ESO 558-G009 is shown in the upper righthand panel of \autoref{fig:hex1}, along with a 22\,GHz spectrum reproduced from \citet{Pesce_2015} (see Fig. 1 in that paper).  The 22\,GHz spectrum is an average of multiple epochs of spectral monitoring observations taken with the GBT.

The only significant component seen in the 183\,GHz spectrum of ESO 558-G009 is the central set of systemic maser features, which span a velocity range of $\sim$40\,\kms around a central velocity of ${\sim}7585$\,\kms.  The peak flux density of the systemic features is $\sim$11.8\,mJy, which is a factor of $\sim$6 fainter than the typical peak flux density of the 22\,GHz systemic masers.

With the overlaid 22\,GHz spectrum in \autoref{fig:hex1}, apparent features in the 183\,GHz spectrum near 7075\,\kms, 7150\,\kms, and 8120\,\kms are suggestive because they coincide with high-velocity features at 22\,GHz.  However, none of these features rise above the $3\sigma$ significance level, so we cannot confirm high-velocity masers in the 183\,GHz spectrum of ESO 558-G009.

Adopting a Hubble law distance of 110\,Mpc, the isotropic luminosity of the 183\,GHz maser emission from ESO 558-G009 is $L_{\text{iso}} \approx 402$\,L$_{\odot}$.  This value is a factor of $\sim$2 lower than the $\sim$709\,L$_{\odot}$ isotropic luminosity observed at 22\,GHz \citep{Pesce_2015}.

\subsection{IC 485}

The 22\,GHz maser system in IC 485 was discovered by the MCP and identified as a disk maser system because of its spectral structure \citep{Pesce_2015}, though the faintness of the high-velocity features makes confirmation of this system's disk nature with VLBI observations difficult (Ladu et al. \textit{in prep.}).  Our 183\,GHz spectrum of IC 485 is shown in the middle lefthand panel of \autoref{fig:hex1}, along with a 22\,GHz spectrum reproduced from \citet{Pesce_2015} (see Fig. 1 in that paper).  The 22\,GHz spectrum is an average of multiple epochs of spectral monitoring observations taken with the GBT.

The only component seen in the 183\,GHz spectrum of IC 485 is the central set of systemic maser features, which span a velocity range of $\sim$60\,\kms around a central velocity of ${\sim}8355$\,\kms.  The peak flux density of the systemic features is $\sim$8.4\,mJy, which is a factor of $\sim$8 fainter than the typical peak flux density of the 22\,GHz systemic masers.  No high-velocity maser features are detected at 183\,GHz, consistent with their relative faintness at 22\,GHz (where they are more than an order of magnitude fainter than the systemic features).

Adopting a Hubble law distance of 119\,Mpc, the isotropic luminosity of the 183\,GHz maser emission from IC 485 is $L_{\text{iso}} \approx 527$\,L$_{\odot}$.  This value is a factor of $\sim$2 lower than the $\sim$1061\,L$_{\odot}$ isotropic luminosity observed at 22\,GHz \citep{Pesce_2015}.

\subsection{J0847-0022}

The 22\,GHz maser system in J0847-0022 was discovered by the MCP and identified as a disk maser system because of its spectral structure \citep{Pesce_2015}, though to date there have been no published VLBI observations to confirm the disk maser nature of this system.  Our 183\,GHz spectrum of J0847-0022 is shown in the middle righthand panel of \autoref{fig:hex1}, along with a 22\,GHz spectrum reproduced from \citet{Pesce_2015} (see Fig. 1 in that paper).  The 22\,GHz spectrum is an average of multiple epochs of spectral monitoring observations taken with the GBT.

The only component seen in the 183\,GHz spectrum of J0847-0022 is the central set of systemic maser features, which span a velocity range of $\sim$70\,\kms around a central velocity of ${\sim}15295$\,\kms.  The peak flux density of the systemic features is $\sim$6.3\,mJy, which is a factor of $\sim$5 fainter than the typical peak flux density of the 22\,GHz systemic masers.  No high-velocity maser features are detected at 183\,GHz despite them being more prominent than the systemic features at 22\,GHz; the high-velocity features must thus be at least an order of magnitude fainter at 183\,GHz than at 22\,GHz.

Adopting a Hubble law distance of 218\,Mpc, the isotropic luminosity of the 183\,GHz maser emission from J0847-0022 is $L_{\text{iso}} \approx 1180$\,L$_{\odot}$.  This value is a factor of $\sim$2.5 lower than the $\sim$2945\,L$_{\odot}$ isotropic luminosity observed at 22\,GHz \citep{Pesce_2015}.

\subsection{Mrk 1419}

The 22\,GHz maser system in Mrk 1419 was discovered by \citet{Henkel_2002} and VLBI observations presented in \citet{Kuo_2011} confirm that it is a disk maser system.  Our 183\,GHz spectrum of Mrk 1419 is shown in the lower lefthand panel of \autoref{fig:hex1}, along with a 22\,GHz spectrum reproduced from \citet{Pesce_2015} (see Fig. 1 in that paper).  The 22\,GHz spectrum is an average of multiple epochs of spectral monitoring observations taken with the GBT.

The only component seen in the 183\,GHz spectrum of Mrk 1419 is the central set of systemic maser features, which span a velocity range of $\sim$100\,\kms around a central velocity of ${\sim}4885$\,\kms.  The peak flux density of the systemic features is $\sim$23.6\,mJy, which is comparable to the typical peak flux density of the 22\,GHz systemic masers.  However, no high-velocity maser features are detected at 183\,GHz despite them being comparably prominent to the systemic features at 22\,GHz; the high-velocity features must thus be at least $\sim$4 times fainter at 183\,GHz than at 22\,GHz.

Adopting a Hubble law distance of 70\,Mpc, the isotropic luminosity of the 183\,GHz maser emission from Mrk 1419 is $L_{\text{iso}} \approx 542$\,L$_{\odot}$.  This value is comparable to the $\sim$565\,L$_{\odot}$ isotropic luminosity observed at 22\,GHz \citep{Pesce_2015}.

\subsection{IC 2560}

The 22\,GHz maser system in IC 2560 was discovered by \citet{Braatz_1996}, and VLBI observations presented in \citet{Nakai_1998} and \citet{Ishihara_2001} confirm that it is a disk maser system.  Our 183\,GHz spectrum of IC 2560 is shown in the lower righthand panel of \autoref{fig:hex1}, along with a 22\,GHz spectrum reproduced from \citet{Pesce_2015} (see Fig. 1 in that paper).  The 22\,GHz spectrum is an average of multiple epochs of spectral monitoring observations taken with the GBT.

The only component seen in the 183\,GHz spectrum of IC 2560 is the central set of systemic maser features, which span a velocity range of $\sim$60\,\kms around a central velocity of ${\sim}2925$\,\kms.  The peak flux density of the systemic features is $\sim$29.6\,mJy, which is a factor of $\sim$5 fainter than the typical peak flux density of the 22\,GHz systemic masers.  The heavily blended structure of the systemic maser complex at 183\,GHz is similar to that at 22\,GHz, with no prominent narrow features present.

Adopting a Hubble law distance of 42\,Mpc, the isotropic luminosity of the 183\,GHz maser emission from IC 2560 is $L_{\text{iso}} \approx 282$\,L$_{\odot}$.  This value is comparable to the $\sim$210\,L$_{\odot}$ isotropic luminosity observed at 22\,GHz \citep{Pesce_2015}.

\subsection{UGC 6093}

\begin{figure*}
    \centering
    \includegraphics[width=1.00\textwidth]{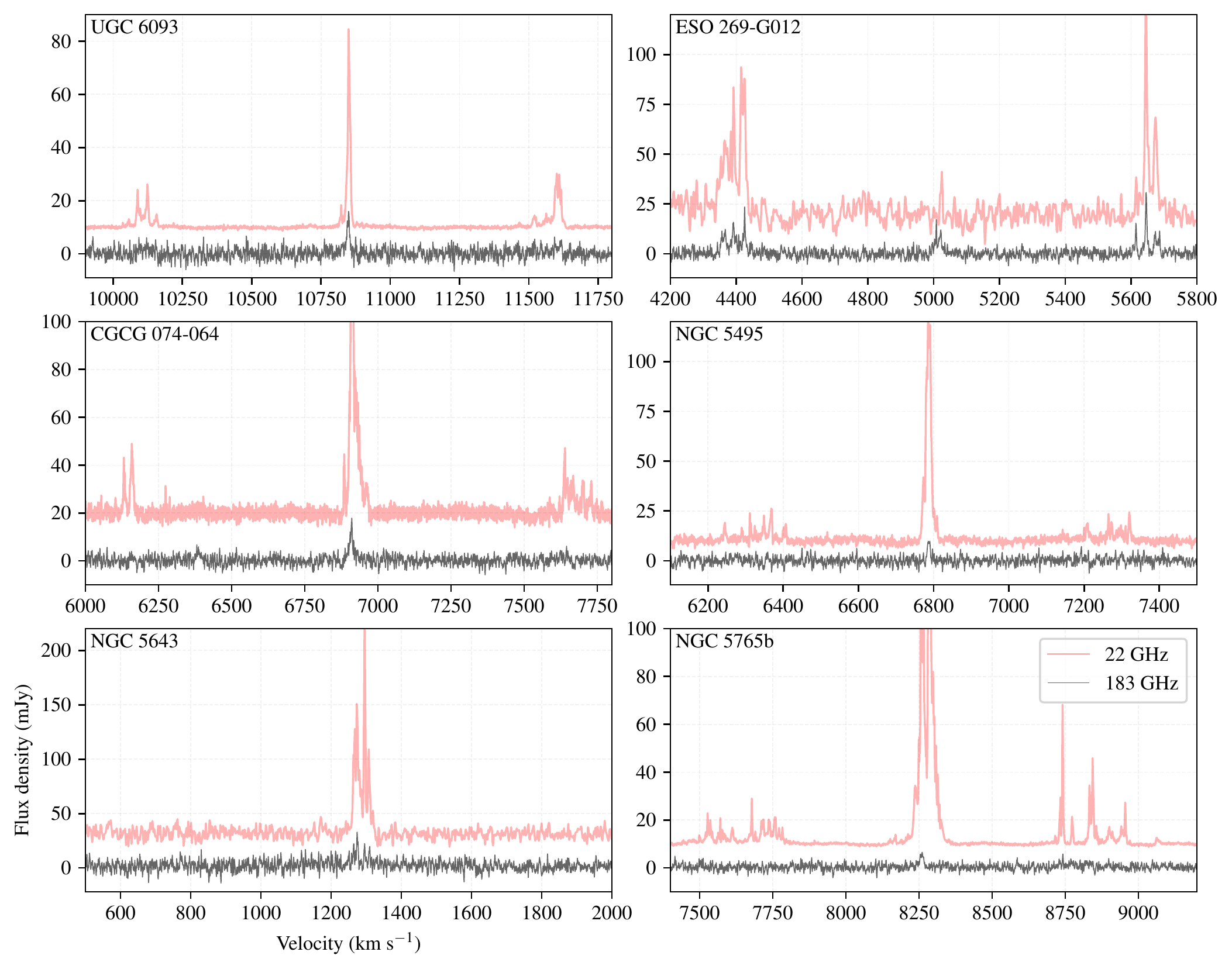}
    \caption{Continuation of \autoref{fig:hex1}.}\label{fig:hex2}
\end{figure*}

The 22\,GHz maser system in UGC 6093 was discovered by the MCP, and VLBI observations presented in \citet{Zhao_2018} confirm that it is a disk maser system.  Our 183\,GHz spectrum of UGC 6093 is shown in the upper lefthand panel of \autoref{fig:hex2}, along with a 22\,GHz spectrum reproduced from \citet{Pesce_2015} (see Fig. 1 in that paper).  The 22\,GHz spectrum is an average of multiple epochs of spectral monitoring observations taken with the GBT.

The only significant component seen in the 183\,GHz spectrum of UGC 6093 is the central set of systemic maser features, which span a velocity range of $\sim$30\,\kms around a central velocity of ${\sim}10848$\,\kms.  The peak flux density of the systemic features is $\sim$16.0\,mJy, which is a factor of $\sim$12 fainter than the typical peak flux density of the 22\,GHz systemic masers.

Similar to ESO 558-G009, the overlaid 22\,GHz spectrum in \autoref{fig:hex2} draws the eye to some features in the 183\,GHz spectrum of UGC 6093 near 10100\,\kms and 11600\,\kms that are suggestive because they coincide with high-velocity features at 22\,GHz.  However, we again find that none of these features rise above the $3\sigma$ significance level, so we cannot confirm the presence of high-velocity masers in the 183\,GHz spectrum of UGC 6093.

Adopting a Hubble law distance of 154\,Mpc, the isotropic luminosity of the 183\,GHz maser emission from UGC 6093 is $L_{\text{iso}} \approx 978$\,L$_{\odot}$.  This value is comparable to the $\sim$1048\,L$_{\odot}$ isotropic luminosity observed at 22\,GHz \citep{Pesce_2015}.

\subsection{ESO 269-G012}

The 22\,GHz maser system in ESO 269-G012 was discovered by \citet{Greenhill_2003b} and identified as a disk maser system because of its spectral structure, though VLBI observations have not been carried out to confirm the disk maser nature of this system.  Our 183\,GHz spectrum of ESO 269-G012 is shown in the upper righthand panel of \autoref{fig:hex2}, along with a 22\,GHz spectrum reproduced from \citet{Greenhill_2003b} (see Fig. 1 in that paper).  The 22\,GHz spectrum was observed with the 70\,m Deep Space Network dish at Tidbinbilla.

The 183\,GHz spectrum of ESO 269-G012 shows three prominent sets of maser features, coinciding in velocity with the three groups of maser features seen in the 22\,GHz spectrum and matching the expected spectral structure for an edge-on disk system.  The line complexes in the 183 GHz spectrum of ESO 269-G012 are relatively rich compared to many of the other spectra in our sample, with each of the three sets of lines containing multiple identifiable maser features.  The peak flux density of the systemic features in the 183\,GHz spectrum is $\sim$17.2\,mJy, which is comparable to the peak flux density of the 22\,GHz systemic masers.  The peak flux density of the high-velocity features is $\sim$30.8\,mJy, which is a factor of $\sim$3 fainter than the peak flux density seen in the 22\,GHz high-velocity masers.

Adopting a Hubble law distance of 72\,Mpc, the isotropic luminosity of the 183\,GHz maser emission from ESO 269-G012 is $L_{\text{iso}} \approx 1337$\,L$_{\odot}$.  This value is roughly 2.5 times larger than the $\sim$496\,L$_{\odot}$ isotropic luminosity observed at 22\,GHz \citep{Greenhill_2003b,Pesce_2015}.

\subsection{CGCG 074-064}

The 22\,GHz maser system in CGCG 074-064 was discovered by the MCP, and VLBI observations presented in \citet{Pesce_2020} confirm that it is a disk maser system.  Our 183\,GHz spectrum of CGCG 074-064 is shown in the middle lefthand panel of \autoref{fig:hex2}, along with a 22\,GHz spectrum reproduced from \citet{Pesce_2015} (see Fig. 1 in that paper).  The 22\,GHz spectrum is an average of multiple epochs of spectral monitoring observations taken with the GBT.

The most prominent component in the 183\,GHz spectrum of CGCG 074-064 is the central set of systemic maser features, which span a velocity range of $\sim$60\,\kms around a central velocity of ${\sim}6910$\,\kms.  The peak flux density of the systemic features is $\sim$17.9\,mJy, which is a factor of $\sim$11 fainter than the typical peak flux density of the 22\,GHz systemic masers.

A second component in the 183\,GHz spectrum of CGCG 074-064 is a weaker (peak flux density of $\sim$6.7\,mJy) feature spanning a velocity range of $\sim$20\,\kms around a central velocity of ${\sim}6387$\,\kms.  This spectral feature coincides with the most redshifted end of the blueshifted complex of high-velocity maser features seen at 22\,GHz.

Adopting the symmetrized $87.6 \pm 7.5$\,Mpc distance measurement from \citet{Pesce_2020}, the isotropic luminosity of the 183\,GHz maser emission from CGCG 074-064 is $L_{\text{iso}} \approx 484$\,L$_{\odot}$.  This value is a factor of $\sim$2 lower than the $\sim$851\,L$_{\odot}$ isotropic luminosity observed at 22\,GHz \citep{Pesce_2015}.

\subsection{NGC 5495}

The 22\,GHz maser system in NGC 5495 was discovered by \citet{Kondratko_2006}, and VLBI observations presented in \citet{Gao_2017} confirm that it is a disk maser system.  Our 183\,GHz spectrum of NGC 5495 is shown in the middle righthand panel of \autoref{fig:hex2}, along with a 22\,GHz spectrum reproduced from \citet{Pesce_2015} (see Fig. 1 in that paper).  The 22\,GHz spectrum is an average of multiple epochs of spectral monitoring observations taken with the GBT.

The only component seen in the 183\,GHz spectrum of NGC 5495 is the central set of systemic maser features, which span a velocity range of $\sim$15\,\kms around a central velocity of ${\sim}6787$\,\kms.  The peak flux density of the systemic features is $\sim$9.9\,mJy, which is a factor of $\sim$12 fainter than the typical peak flux density of the 22\,GHz systemic masers.

Adopting a Hubble law distance of 96\,Mpc, the isotropic luminosity of the 183\,GHz maser emission from NGC 5495 is $L_{\text{iso}} \approx 163$\,L$_{\odot}$.  This value is a factor of $\sim$4 lower than the $\sim$625\,L$_{\odot}$ isotropic luminosity observed at 22\,GHz \citep{Pesce_2015}.

\subsection{NGC 5643}

The 22\,GHz maser system in NGC 5643 was discovered by \citet{Greenhill_2003b}.  The 22\,GHz spectrum contains only systemic features, and thus may not be associated with an edge-on disk; VLBI observations have also not yet been carried out on this system.  Our 183\,GHz spectrum of NGC 5643 is shown in the lower lefthand panel of \autoref{fig:hex2}, along with a 22\,GHz spectrum reproduced from \citet{Greenhill_2003b} (see Fig. 1 in that paper).  The 22\,GHz spectrum was observed with the 70\,m Deep Space Network dish at Tidbinbilla.

The only component seen in the 183\,GHz spectrum of NGC 5643 is the central set of systemic maser features, which span a velocity range of $\sim$90\,\kms around a central velocity of ${\sim}1275$\,\kms.  The peak flux density of the systemic features is $\sim$32.9\,mJy, which is a factor of $\sim$6 fainter than the typical peak flux density of the 22\,GHz systemic masers.

Because NGC 5643 is nearby enough that the Hubble law distance is expected to be very uncertain, we instead adopt a Type Ia Supernova distance of $12.4 \pm 0.5$\,Mpc to this system from \citet{Burns_2020}.  The isotropic luminosity of the 183\,GHz maser emission from NGC 5643 is $L_{\text{iso}} \approx 27$\,L$_{\odot}$.  This value is a factor of $\sim$2 higher than the $\sim$12.6\,L$_{\odot}$ isotropic luminosity observed at 22\,GHz \citep{Greenhill_2003b}.

\subsection{NGC 5765b}

The 22\,GHz maser system in NGC 5765b was discovered by the MCP, and VLBI observations presented in \citet{Gao_2016} confirm that it is a disk maser system.  Our 183\,GHz spectrum of NGC 5765b is shown in the bottom righthand panel of \autoref{fig:hex2}, along with a 22\,GHz spectrum reproduced from \citet{Pesce_2015} (see Fig. 1 in that paper).  The 22\,GHz spectrum is an average of multiple epochs of spectral monitoring observations taken with the GBT.

The most prominent component seen in the 183\,GHz spectrum of NGC 5765b is the central set of systemic maser features, which span a velocity range of $\sim$20\,\kms around a central velocity of ${\sim}8260$\,\kms.  The peak flux density of the systemic features is $\sim$6.5\,mJy, which is a factor of $\sim$30 fainter than the typical peak flux density of the 22\,GHz systemic masers.

A potential second component near $\sim$8740\,\kms is most apparent in the zoomed-in spectrum shown in \autoref{fig:spectra}, and is coincident with the brightest redshifted maser features seen in the 22\,GHz spectrum (see \autoref{fig:hex2}).  However, this feature does not rise above the $3\sigma$ significance level, so we cannot confirm the presence of high-velocity masers in the 183\,GHz spectrum of NGC 5765b.

Adopting the $126.3 \pm 11.6$\,Mpc distance from \citet{Gao_2016}, the isotropic luminosity of the 183\,GHz maser emission from NGC 5765b is $L_{\text{iso}} \approx 379$\,L$_{\odot}$.  This value is a factor of $\sim$7 lower than the $\sim$2553\,L$_{\odot}$ isotropic luminosity observed at 22\,GHz \citep{Pesce_2015}.

\subsection{Circinus} \label{sec:Circinus}

\begin{figure*}
    \centering
    \includegraphics[width=1.00\textwidth]{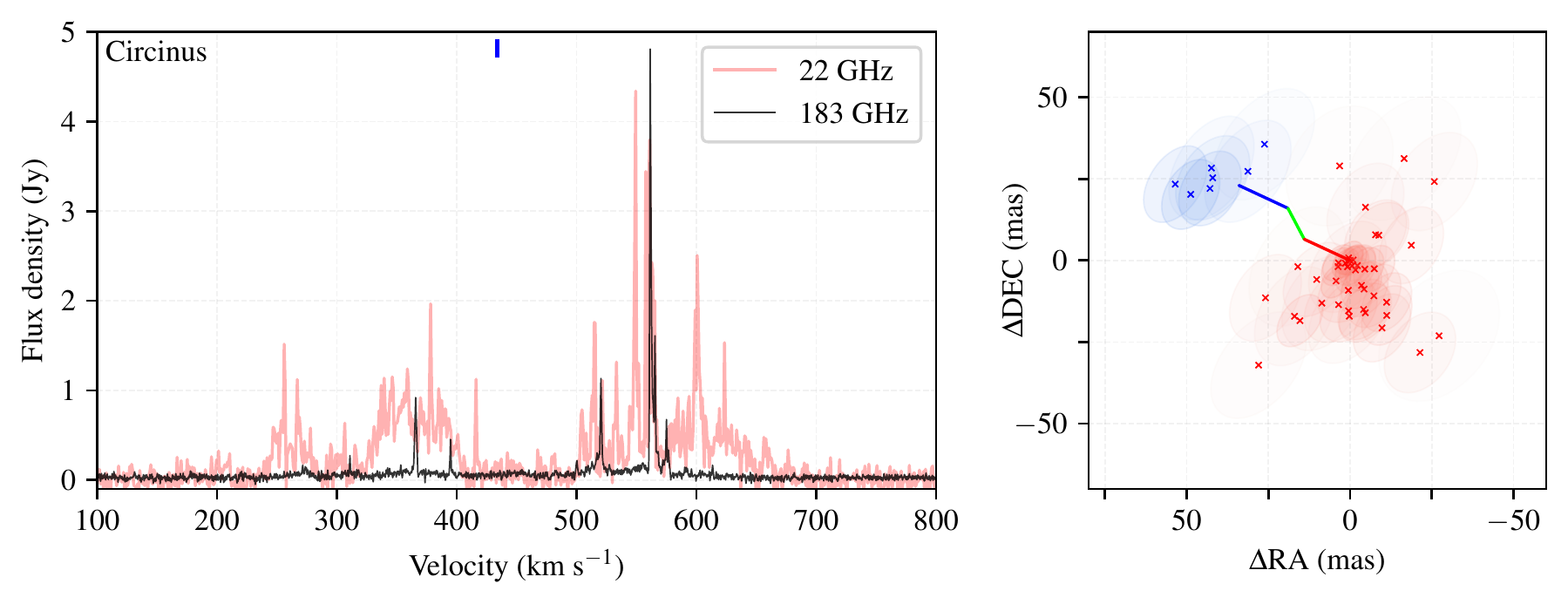}
    \caption{Spectra (left) and map (right) for the Circinus galaxy.  The left panel shows a comparison of the 183\,GHz spectrum (in black) and the 22\,GHz spectrum (in red), as in \autoref{fig:hex1} and \autoref{fig:hex2}; the 22\,GHz spectrum was taken with the Parkes Telescope in 1997 August and is reproduced from \citet{Braatz_2003} (see Fig. 1 in that paper).  The systemic velocity of the galaxy is marked by the blue line at the top.  The right panel shows the best-fit positions of the detected 183\,GHz maser spots as cross markers along with their corresponding $1\sigma$ error ellipses; we show only those masers with mean position uncertainties less than 20\,mas.  Each maser spot in the right panel is colored blue if it has a velocity that is blueshifted with respect to the recession velocity of the galaxy and red if it has a velocity that is redshifted with respect to the recession velocity of the galaxy.  The opacity of the error ellipses scales the signal-to-noise ratio of the fit, with more tightly-constrained positions having a more opaque ellipse.  The overplotted blue, green, and red lines in the right panel mark the warped edge-on disk from \cite{Greenhill_2003} (see Fig. 3 in that paper).}\label{fig:Circinus}
\end{figure*}

The 22\,GHz maser system in Circinus was only the second ever discovered in an AGN \citep{Gardner_1982}, but because of its very southern location (declination of $-65$ degrees) it was not mapped with VLBI observations until the work of \citet{Greenhill_2003}.  The VLBI observations reveal that the masers in this system originate from both a warped, edge-on accretion disk (within the innermost $\sim$0.4\,pc) as well as a wide-angle outflow extending out to $\sim$1\,pc from the AGN, with the entire complex covering an angular region ${\sim}90\text{\,mas} \times 50$\,mas on the sky.  \citet{Greenhill_2003} showed that the masers associated with the disk and those associated with the outflow largely span different velocity ranges, with the latter primarily emitting in the range $\sim$300--570\,\kms and the former primarily emitting outside of that range.  Our 183\,GHz spectrum of Circinus is shown in the lefthand panel of \autoref{fig:Circinus}, along with a 22\,GHz spectrum reproduced from \citet{Braatz_2003}.  The 22\,GHz spectrum was observed with the 64\,m Parkes Telescope in 1997 August.

The 183\,GHz spectrum of Circinus is by far the brightest in our sample; its peak flux density of nearly 5\,Jy is more than two orders of magnitude brighter than that of the next-brightest source, and it is comparable to the peak flux density of the 22\,GHz spectrum from \citet{Braatz_2003}.  However, we note that the 22\,GHz spectrum is known to be highly time-variable \citep{Greenhill_1997,McCallum_2007}, and it has been observed with peak flux densities as high as $\sim$40\,Jy \citep[e.g.,][]{Greenhill_2003}.  The 183\,GHz maser complex in Circinus is characterized by a broad pedestal of emission at a level of $\sim$50--100\,mJy and spanning a velocity range from $\sim$250--650\,\kms, interspersed with a number of bright (flux density $\gtrsim$1\,Jy), narrow ($\text{FWHM} \lesssim 1$\,\kms) individual maser lines.  The velocity range covered by the 183\,GHz maser lines overlaps primarily with the velocity range associated with the outflow masers at 22\,GHz \citep{Greenhill_2003}.

Because Circinus is nearby enough that the Hubble law distance is very uncertain, we instead adopt the Tully-Fisher distance measurement of $4.2 \pm 0.8$\,Mpc to this system from \citet{Karachentsev_2013}.  The isotropic luminosity of the 183\,GHz maser emission from Circinus is $L_{\text{iso}} \approx 171$\,L$_{\odot}$.  This value is a factor of $\sim$2.5 higher than the $\sim$73\,L$_{\odot}$ isotropic luminosity observed at 22\,GHz \citep[e.g.,][]{Braatz_2003}.

Although ALMA was in a relatively compact (C-4) configuration for the 2018 December observations of Circinus (with a corresponding resolution of $\sim$0.6\,arcseconds at 183\,GHz), the high signal-to-noise ratio provided by the bright maser lines enables coarse mapping of the 183\,GHz system.  When fitting an emission structure observed with some width $\Delta$ and signal-to-noise ratio $\rho$, the expected uncertainty in the recovered position of the structure is proportional to $\Delta / \rho$ \citep[see, e.g.,][]{Kaper_1966,Condon_1997}.  With a $\sim$0.6\,arcsecond beam and signal-to-noise ratios exceeding 200 for some lines, we expect to achieve relative positional uncertainties of several mas for a number of the maser features in the 183\,GHz spectrum.  To map the 183\,GHz maser system in Circinus, we fit a point source model directly to the visibilities in each spectral channel individually.  Our fitting procedure is detailed in \autoref{app:PositionModeling}, and the resulting map is shown in the righthand panel of \autoref{fig:Circinus}.  We plot only the maser features that have geometric mean fitted positional uncertainties less than 20\,mas; i.e., we require that $\sqrt{\sigma_x \sigma_y} < 20$\,mas, given an uncertainty in the $x$-position of $\sigma_x$ and an uncertainty in the $y$-position of $\sigma_y$.

Our map shows that the 183\,GHz maser system in Circinus is spatially resolved by these ALMA observations, covering a region that appears to be well-matched in size and shape to the ${\sim}90\text{\,mas} \times 50$\,mas distribution of 22\,GHz masers.  We have colored each plotted maser spot in \autoref{fig:Circinus} by whether it is redshifted or blueshifted with respect to the 434\,\kms systemic velocity of the galaxy.  We find that the orientation and sense of rotation for the 183\,GHz system also qualitatively matches that seen at 22\,GHz, with redshifted material located towards the southwest and blueshifted material located towards the northeast.

To compare more directly with the 22\,GHz maser distribution, we overplot in \autoref{fig:Circinus} the midline of the edge-on disk inferred in \cite{Greenhill_2003}.  We note that \cite{Greenhill_2003} referenced their VLBI map to a maser spot with a velocity\footnote{\cite{Greenhill_2003} quote the velocity of their reference spot as being 565.2\,\kms using the radio convention in the heliocentric reference frame.  We have converted the velocity here to the optical convention, as is used throughout the rest of this paper.} of 566.3\,\kms, while we have referenced our map to a maser spot with a velocity of 561.6\,\kms; in both cases, the reference maser is the brightest one in the spectrum.  It is plausible that both of these lines originate from the same parcel of masing gas (with the $\sim$5\,\kms difference in velocity then attributable to dynamical evolution during the $\sim$20\,years separating the two observations), but the lack of absolute astrometric referencing between the two observations means that we cannot be sure; the relative locations of the 183\,GHz maser system and the overplotted 22\,GHz disk midline in \autoref{fig:Circinus} should thus be understood to be an assumption.  Proceeding with this assumption, we find that the 183\,GHz masers live primarily outside of the disk region, with a $\sim$40\,mas gap separating the blueshifted and redshifted complexes.

The 183\,GHz masers are consistent with being located in the disk identified at 22\,GHz, but at larger orbital radii than are occupied by the 22\,GHz disk masers.  Having 183\,GHz masers probe out to larger orbital radii than their 22\,GHz counterparts would also be consistent with the results of radiative transfer modeling, in which the 183\,GHz transition is optimally pumped at lower gas densities and temperatures than the 22\,GHz transition \citep{Yates_1997,Gray_2016}.  However, the current large uncertainties in our maser position measurements (relative to the size of the system) ultimately make it difficult to unambiguously distinguish between a disk and outflow origin for the 183\,GHz masers in Circinus.

\subsection{Nondetections} \label{sec:Nondetections}

The remaining 7 of the 20 systems targeted in our survey are not detected in 183\,GHz maser emission.  Of these nondetections, 5 of them are disk systems known to host relatively weak (${\lesssim}$60\,mJy peak flux density) 22\,GHz maser systems \citep{Pesce_2015}.  If the 183\,GHz emission in these systems is an order of magnitude fainter than the corresponding 22\,GHz emission -- as is the case for a number of our detections -- then our 183\,GHz observations would simply not be sensitive enough to detect it.

The other two nondetections -- NGC 1068 and NGC 1386 -- host strong ($\sim$1\,Jy peak flux density) 22\,GHz maser systems \citep{Claussen_1984,Braatz_1996}, and we would have strongly detected 183\,GHz emission from these systems even if it were an order of magnitude fainter than the corresponding 22\,GHz emission.  It is thus perhaps surprising that these systems are not detected by our 183\,GHz observations, because the nondetection implies that any 183\,GHz emission that might be present is substantially more than an order of magnitude fainter than the corresponding 22\,GHz emission.  It is also interesting to note that the 22\,GHz maser systems in both NGC 1068 and NGC 1386 exhibit more complicated dynamics and geometries than many of the more ordered edge-on disk systems in our survey.  The 22\,GHz maser system in NGC 1068 is distributed in an approximately linear structure on the sky \citep{Greenhill_1996}, but it exhibits sub-Keplerian rotation velocities and may trace the upper layers of a geometrically thick torus or disk structure whose surface is irradiated by the jet \citep[e.g.,][]{Gallimore_2001}; the 22\,GHz maser system in NGC 1386 is not consistent with a rotating structure at all \citep{Braatz_1997}.  The lack of observed 183\,GHz maser emission from these systems may thus be associated with their distinct underlying physical conditions compared with most of the other systems in our survey.

\section{Summary and conclusions} \label{sec:Summary}

In this paper we present the results of a survey to identify 183\,GHz water maser emission from AGN targets that are known to host 22\,GHz megamaser emission.  We used ALMA to observe 20 targets with a typical spectral sensitivity of $\sim$2\,mJy per $\sim$2\,\kms of bandwidth, and we detect 183\,GHz maser emission in 13 of these targets.  In those targets for which both 183\,GHz continuum and spectral line emission is detected, we find that the spectral line emission is spatially coincident with the peak of the continuum emission.

For the majority (9 out of 13) of our 183\,GHz maser detections, the spectrum exhibits only a single, largely featureless line complex located near the systemic velocity of the galaxy.  Of the remaining 4 sources, two (NGC 1194 and CGCG 074-064) have a single additional spectral feature detected at velocities that are offset by several hundred \kms from the systemic velocity of the galaxy, while the other two (Circinus and ESO 269-G012) exhibit considerably more complex spectral structure.  Across all targets, we find that the 183\,GHz maser emission is systematically fainter than the corresponding 22\,GHz emission from the same system (typically by nearly an order of magnitude in flux density).  However, because the emission frequency of the 183\,GHz transition is also approximately an order of magnitude higher than that of the 22\,GHz transition, the isotropic luminosities at 183\,GHz and 22\,GHz are comparable (typically within a factor of $\sim$2 of one another).  I.e., assuming similar maser beaming angles, the energy emitted from the 183\,GHz transition is similar to that from the 22\,GHz transition.

The 183\,GHz spectra of Circinus and ESO 269-G012 contain the richest line complexes in our sample, with the latter clearly exhibiting the triple-peaked spectral structure characteristic of edge-on disk systems seen at 22\,GHz \citep{Pesce_2015}.  The qualitative difference in spectral complexity between these two sources and the others in our sample likely arises from their higher signal-to-noise ratio relative to most of the other spectra in our sample; the situation is reminiscent of the increased spectral richness observed in 22\,GHz maser systems with the GBT compared to prior observations with less sensitive facilities \citep{Braatz_2003}.

Among the galaxies targeted in our sample, Circinus exhibits the strongest 183\,GHz emission by more than two orders of magnitude in flux density, with the brightest line peaking at nearly 5\,Jy.  Even though the nominal beam size of these ALMA observations is arcsecond-scale, the high signal-to-noise ratios achieved by the strong lines in the Circinus spectrum enable us to carry out coarse mapping of the mas-scale system.  We find that the spatial distribution of the 183\,GHz masers is qualitatively similar to that of the 22\,GHz masers (which were mapped with VLBI by \citealt{Greenhill_2003}), with the blueshifted masers located to the northeast of the redshifted masers and the entire system covering ${\sim}90\text{\,mas} \times 50$\,mas on the sky.  The current observations are most consistent with the 183\,GHz material originating from the disk identified in the 22\,GHz system, but at larger orbital radii than are occupied by the 22\,GHz disk masers.

The survey presented in this paper demonstrates the existence of a population of 183\,GHz AGN megamasers that shows every indication of being associated with the same edge-on accretion disk material as the known 22\,GHz population.  We thus expect that VLBI observations of these systems at 183\,GHz -- which we note are not currently possible with any existing facility -- would reveal simple orbital dynamics that present the same utility for measurements of distances and SMBH masses as the 22\,GHz systems currently do, but with roughly an order of magnitude finer angular resolution.  In the meantime, we have demonstrated that even connected-element observations with ALMA are capable of coarse mapping of the strongest 183\,GHz systems.  Extended-configuration ALMA observations of the Circinus galaxy should permit mapping of the 183\,GHz maser system with an effective resolution that is comparable to that of the best current 22\,GHz maps.

\facilities{ALMA}

\software{\texttt{astropy} \citep{Astropy_2013,Astropy_2018,Astropy_2022}, CASA \citep{McMullin_2007}, \texttt{dynesty} \citep{Speagle_2020}, \texttt{matplotlib} \citep{Hunter_2007}, \texttt{numpy} \citep{Harris_2020}, \texttt{scipy} \citep{Virtanen_2020}}

\acknowledgments

Support for this work was provided by the NSF through grant AST-1935980, and by the Gordon and Betty Moore Foundation through grant GBMF-5278.  This work has been supported in part by the Black Hole Initiative at Harvard University, which is funded by grants from the John Templeton Foundation and the Gordon and Betty Moore Foundation to Harvard University.  This paper makes use of the following ALMA data: ADS/JAO.ALMA\#2018.1.00321.S, ADS/JAO.ALMA\#2017.1.00909.S. ALMA is a partnership of ESO (representing its member states), NSF (USA) and NINS (Japan), together with NRC (Canada), MOST and ASIAA (Taiwan), and KASI (Republic of Korea), in cooperation with the Republic of Chile. The Joint ALMA Observatory is operated by ESO, AUI/NRAO and NAOJ.  The National Radio Astronomy Observatory is a facility of the National Science Foundation operated under cooperative agreement by Associated Universities, Inc.  This research has made use of the NASA/IPAC Extragalactic Database (NED), which is operated by the Jet Propulsion Laboratory, California Institute of Technology, under contract with the National Aeronautics and Space Administration.

\bibliography{references}{}
\bibliographystyle{aasjournal}

\appendix
\numberwithin{equation}{section}

\section{Additional calibration details} \label{app:Calibration}

In this section we provide specific calibration details relevant for individual science targets.  We focus primarily on the flux density calibration fidelity, which is unique to each target, though we also highlight any elements of the calibration that deviate from the standard procedure described in \autoref{sec:Calibration}.  The calibrator details for each science target are listed in \autoref{tab:Calibrators}.

The flux density calibrators contained in the ALMA catalog typically exhibit substantial time variability (i.e., varying by factors of several in total flux density over the multi-year monitoring period), so when applying the flux density calibration we pick the dates that fall closest in time to the observations for each source.  The ALMA monitoring observations typically do not have observations at 183\,GHz directly, so we instead interpolate the observed flux densities -- one higher frequency $\nu_{\text{high}}$ and one lower frequency $\nu_{\text{low}}$ -- to determine the appropriate 183\,GHz calibration.

We also use the historical ALMA monitoring observations to empirically estimate the reliability of the flux density calibration for each target.  For all monitoring epochs in which a third frequency $\nu_{\text{mid}}$ that falls between $\nu_{\text{low}}$ and $\nu_{\text{high}}$ was also observed, we take a difference between the observed flux density at $\nu_{\text{mid}}$ and that predicted by the spectral index derived from the observations at $\nu_{\text{low}}$ and $\nu_{\text{high}}$.  We take the standard deviation of these differences over all historical ALMA monitoring observations as our measure of flux density calibration fidelity, which we add in quadrature to the statistical uncertainty and then cast as a fraction $f_{\text{cal}}$ of the flux density itself; this value is reported in \autoref{tab:Calibrators} for each target.  We find a typical flux density precision between 5\% and 10\%.

\begin{deluxetable*}{lccccccc}
\tablecolumns{8}
\tablewidth{0pt}
\tablecaption{Calibrator details\label{tab:Calibrators}}
\tablehead{ & & & \multicolumn{5}{c}{Flux density calibrator details} \\\cline{4-8}
\colhead{Science target} & \colhead{Bandpass calibrator} & \colhead{Gain calibrator}  & \colhead{Calibrator} & \colhead{$S_{\nu,0}$ (Jy)} & \colhead{$\nu_0$ (GHz)} & \colhead{$\alpha$} & \colhead{$f_{\text{cal}}$ (\%)}}
\startdata
J0109-0332 & J0006-0623 & J0108+0135 & J0006-0623 & 1.98 & 103.5 & $-0.687$ & 8.0 \\
J0126-0417 & J0006-0623 & J0141-0928 & J0006-0623 & 1.98 & 103.5 & $-0.687$ & 8.0 \\
NGC 1068 (2018 Nov 09) & J0423-0120 & J0217+0144 & J0423-0120 & 3.59 & 103.5 & $-0.496$ & 9.0 \\
NGC 1068 (2018 Nov 16) & J0423-0120 & J0217+0144 & J0423-0120 & 3.59 & 103.5 & $-0.496$ & 9.0 \\
NGC 1194 & J0238+1636 & J0241-0815 & J0238+1636 & 0.59 & 103.5 & $-0.653$ & 5.7 \\
NGC 1386 & J0522-3627 & J0403-3605 & J0522-3627 & 5.46 & 103.5 & $-0.132$ & 4.1 \\
ESO 558-G009 & J0522-3627 & J0702-1951 & J0522-3627 & 5.35 & 103.5 & $-0.230$ & 4.1 \\
IC 485 (2018 Oct 31) & J0854+2006 & J0748+2400 & J0854+2006 & 4.51 & 103.5 & $-0.337$ & 4.2 \\
IC 485 (2018 Dec 06) & J0854+2006 & J0748+2400 & J0854+2006 & 4.52 & 103.5 & $-0.487$ & 4.2 \\
J0847-0022 & J0750+1231 & J0839+0104 & J0750+1231 & 0.97 & 103.5 & $-0.643$ & 8.7 \\
Mrk 1419 & J0854+2006 & J0930+0034 & J0854+2006 & 4.08 & 103.5 & $-0.496$ & 4.3 \\
IC 2560 (2018 Oct 02) & J1058+0133 & J1037-2934 & J1058+0133 & 4.10 & 91.5 & $-0.488$ & 5.2 \\
IC 2560 (2018 Oct 30) & J1058+0133 & J1037-2934 & J1058+0133 & 4.06 & 103.5 & $-0.542$ & 4.0 \\
IC 2560 (2018 Nov 08) & J1107-4449 & J1037-2934 & J1107-4449 & 1.37 & 103.5 & $-0.677$ & 6.6 \\
NGC 3393 (2018 Oct 30) & J1107-4449 & J1037-2934 & J1107-4449 & 1.36 & 103.5 & $-0.712$ & 6.1 \\
NGC 3393 (2018 Nov 14) & J1107-4449 & J1037-2934 & J1107-4449 & 1.41 & 103.5 & $-0.701$ & 6.6 \\
UGC 6093 & J1058+0133 & J1041+0610 & J1058+0133 & 4.06 & 103.5 & $-0.542$ & 4.0 \\
ESO 269-G012 & J1427-4206 & J1326-5256 & J1427-4206 & 3.77 & 91.5 & $-0.585$ & 6.2 \\
CGCG 074-064 & J1256-0547 & J1415+1320 & J1256-0547 & 14.93 & 91.5 & $-0.537$ & 6.5 \\
NGC 5495 & J1517-2422 & J1408-2900 & J1517-2422 & 3.15 & 91.5 & $-0.321$ & 5.7 \\
Circinus & J1427-4206 & J1326-5256 & J1427-4206 & 3.77 & 91.5 & $-0.585$ & 6.2 \\
NGC 5643 (2018 Aug 27) & J1256-0547 & J1427-4206 & J1256-0547 & 11.42 & 103.5 & $-0.548$ & 6.4 \\
NGC 5643 (2018 Sep 20) & J1517-2422 & J1427-4206 & J1517-2422 & 2.84 & 103.5 & $-0.256$ & 5.4 \\
NGC 5765b & J1504+1029 & J1504+1029 & J1504+1029 & 1.96 & 103.5 & $-0.626$ & 4$^*$ \\
CGCG 165-035 & J1550+0527 & J1516+1932 & J1550+0527 & 0.90 & 91.5 & $-0.809$ & 5.7 \\
NGC 6264 & J1751+0939 & J1653+3107 & J1751+0939 & 3.57 & 103.5 & $-0.451$ & 4.4 \\
\enddata
\tablecomments{Information about the calibrator targets used for each of the observations reported in this paper.\\
$^*$Because we use the targeted gain calibrator rather than the targeted flux calibrator for NGC 5765b flux density calibration, there are insufficient historical ALMA observations to estimate the flux density calibration precision.  The $f_{\text{cal}}$ value reported here thus reflects just the statistical uncertainty in the observations at $\nu_{\text{low}}$ and $\nu_{\text{high}}$.}
\end{deluxetable*}

\subsection{J0109-0332}

The flux density calibrator for J0109-0332 is J0006-0623.  Its flux density as reported by ALMA on 2018 Nov 08 was $0.88 \pm 0.04$\,Jy at 337.5\,GHz, and on 2018 Nov 09 it was $1.98 \pm 0.04$\,Jy at 103.5\,GHz.  We use these values to derive a spectral index of $\alpha = -0.687 \pm 0.042$.

\subsection{J0126-0417}

The flux density calibrator for J0126-0417 is J0006-0623, which is the same as for J0109-0332.

\subsection{NGC 1068}

NGC 1068 was observed twice, once on 2018 Nov 09 and again on 2018 Nov 16.  The flux density calibrator for NGC 1068 was J0423-0120 during both observations.  Its flux density as reported by ALMA on 2018 Nov 10 was $1.98 \pm 0.05$\,Jy at 343.5\,GHz, and on 2018 Nov 12 it was $3.59 \pm 0.07$\,Jy at 103.5\,GHz.  We use these values to derive a spectral index of $\alpha = -0.496 \pm 0.027$.

Because NGC 1068 was observed on two separate days, we combine the MSs using CASA's \texttt{concat} task after calibration and prior to imaging.

\subsection{NGC 1194}

The flux density calibrator for NGC 1194 is J0238+1636.  Its flux density as reported by ALMA on 2018 Nov 19 was $0.27 \pm 0.02$\,Jy at 343.5\,GHz and $0.59 \pm 0.03$\,Jy at 103.5\,GHz, which we use to derive a spectral index of $\alpha = -0.653 \pm 0.076$.

\subsection{NGC 1386}

The flux density calibrator for NGC 1386 is J0522-3627.  Its flux density as reported by ALMA on 2018 Nov 19 was $4.65 \pm 0.10$\,Jy at 349.5\,GHz and $5.46 \pm 0.14$\,Jy at 103.5\,GHz, which we use to derive a spectral index of $\alpha = -0.132 \pm 0.028$.

\subsection{ESO 558-G009}

The flux density calibrator for ESO 558-G009 is J0522-3627.  Its flux density as reported by ALMA on 2018 Oct 30 was $4.06 \pm 0.09$\,Jy at 343.5\,GHz and $5.35 \pm 0.14$\,Jy at 103.5\,GHz, which we use to derive a spectral index of $\alpha = -0.230 \pm 0.029$.

\subsection{IC 485}

IC 485 was observed twice, once on 2018 Oct 31 and again on 2018 Dec 06.  The flux density calibrator for IC 485 was J0854+2006 during both observations.  Its flux density as reported by ALMA on 2018 Oct 30 was $3.01 \pm 0.06$\,Jy at 343.5\,GHz and $4.51 \pm 0.07$\,Jy at 103.5\,GHz.  We use these values to derive a spectral index of $\alpha = -0.337 \pm 0.021$.

The flux density of J0854+2006 as reported by ALMA on 2018 Dec 01 was $2.52 \pm 0.05$\,Jy at 343.5\,GHz, and on 2018 Dec 02 it was $4.52 \pm 0.07$\,Jy at 103.5\,GHz.  We use these values to derive a spectral index of $\alpha = -0.487 \pm 0.021$.

Because IC 485 was observed on two separate days, we combine the MSs using CASA's \texttt{concat} task after calibration and prior to imaging.

\subsection{J0847-0022}

The flux density calibrator for J0847-0022 is J0750+1231.  Its flux density as reported by ALMA on 2018 Nov 10 was $0.97 \pm 0.03$\,Jy at 103.5\,GHz and $0.45 \pm 0.04$\,Jy at 343.5\,GHz, which we use to derive a spectral index of $\alpha = -0.643 \pm 0.079$.

\subsection{Mrk 1419}

The flux density calibrator for Mrk 1419 is J0854+2006.  Its flux density as reported by ALMA on 2018 Nov 9 was $4.08 \pm 0.09$\,Jy at 103.5\,GHz and $2.25 \pm 0.06$\,Jy at 343.5\,GHz, which we use to derive a spectral index of $\alpha = -0.496 \pm 0.029$.

\subsection{IC 2560}

IC 2560 was observed three times, on 2018 Oct 02, 2018 Oct 30, and 2018 Nov 08.  The flux density calibrator for IC 2560 was J1058+0133 during the first two observations and J1107-4449 during the third observation.

The flux density of J1058+0133 as reported by ALMA on 2018 Sep 27 was $4.10 \pm 0.10$\,Jy at 91.5\,GHz, and on 2018 Oct 02 it was $2.15 \pm 0.05$\,Jy at 343.5\,GHz.  We use these values to derive a spectral index of $\alpha = -0.488 \pm 0.025$ for the 2018 Oct 02 observation.

The flux density of J1058+0133 as reported by ALMA on 2018 Oct 30 was $4.06 \pm 0.08$\,Jy at 103.5\,GHz and $2.12 \pm 0.06$\,Jy at 343.5\,GHz.  We use these values to derive a spectral index of $\alpha = -0.542 \pm 0.029$ for the 2018 Oct 30 observation.

The flux density of J1107-4449 as reported by ALMA on 2018 Nov 10 was $1.37 \pm 0.04$\,Jy at 103.5\,GHz and $0.61 \pm 0.05$\,Jy at 343.5\,GHz.  We use these values to derive a spectral index of $\alpha = -0.677 \pm 0.073$ for the 2018 Nov 08 observation.

Because IC 2560 was observed on three separate days, we combine the MSs using CASA's \texttt{concat} task after calibration and prior to imaging.

\subsection{NGC 3393}

NGC 3393 was observed twice, once on 2018 Oct 30 and again on 2018 Nov 14.  The flux density calibrator for NGC 3393 was J1107-4449 during both observations.  Its flux density as reported by ALMA on 2018 Oct 30 was $1.36 \pm 0.03$\,Jy at 103.5\,GHz and $0.58 \pm 0.04$\,Jy at 343.5\,GHz, which we use to derive a spectral index of $\alpha = -0.712 \pm 0.061$ for the 2018 Oct 30 observation.  Its flux density as reported by ALMA on 2018 Nov 10 was $1.41 \pm 0.04$\,Jy at 103.5\,GHz and $0.61 \pm 0.05$\,Jy at 343.5\,GHz, which we use to derive a spectral index of $\alpha = -0.701 \pm 0.073$ for the 2018 Nov 14 observation.

Because NGC 3393 was observed on two separate days, we combine the MSs using CASA's \texttt{concat} task after calibration and prior to imaging.

\subsection{UGC 6093}

The flux density calibrator for UGC 6093 is J1058+0133.  Its flux density as reported by ALMA on 2018 Oct 30 was $4.06 \pm 0.08$\,Jy at 103.5\,GHz and $2.12 \pm 0.06$\,Jy at 343.5\,GHz, which we use to derive a spectral index of $\alpha = -0.542 \pm 0.029$.

\subsection{ESO 269-G012}

The flux density calibrator for ESO 269-G012 is J1427-4206.  Its flux density as reported by ALMA on 2018 Dec 01 was $1.74 \pm 0.05$\,Jy at 343.5\,GHz, and on 2018 Dec 02 it was $3.77 \pm 0.08$\,Jy at 91.5\,GHz.  We use these values to derive a spectral index of $\alpha = -0.585 \pm 0.027$.

\subsection{CGCG 074-064}

The flux density calibrator for CGCG 074-064 is J1256-0547.  Its flux density as reported by ALMA on 2018 Dec 01 was $7.34 \pm 0.16$\,Jy at 343.5\,GHz, and on 2018 Dec 02 it was $14.93 \pm 0.29$\,Jy at 91.5\,GHz.  We use these values to derive a spectral index of $\alpha = -0.537 \pm 0.022$.

\subsection{NGC 5495}

The flux density calibrator for NGC 5495 is J1517-2422.  Its flux density as reported by ALMA on 2018 Dec 06 was $2.06 \pm 0.07$\,Jy at 343.5\,GHz, and on 2018 Dec 08 it was $3.15 \pm 0.07$\,Jy at 91.5\,GHz.  We use these values to derive a spectral index of $\alpha = -0.321 \pm 0.031$.

\subsection{Circinus} \label{app:Circinus}

The flux density calibrator for Circinus is J1427-4206.  Its flux density as reported by ALMA on 2018 Dec 01 was $1.74 \pm 0.05$\,Jy at 343.5\,GHz, and on 2018 Dec 02 it was $3.77 \pm 0.08$\,Jy at 91.5\,GHz.  We use these values to derive a spectral index of $\alpha = -0.585 \pm 0.027$.

After initial gain calibration, we further self-calibrate the phases to the strongest spectral line, averaging over the three channels ($\sim$0.6\,\kms) centered at 561.6\,\kms and using a solution interval of 60 seconds.  These phase solutions are applied to all four spectral windows prior to imaging.

\subsection{NGC 5643}

Note that the observing setup for NGC 5643 is different than for the other targets.  It was observed over two days -- 2018 Aug 27 and 2018 Sep 20 -- and its spectral window configuration did not contain a dedicated continuum spectral window.  The flux density calibrator for NGC 5643 was J1256-0547 during the first observation and J1517-2422 during the second observation.

The flux density of J1256-0547 as reported by ALMA on 2018 Aug 28 was $11.42 \pm 0.19$\,Jy at 103.5\,GHz and $5.92 \pm 0.14$\,Jy at 343.5\,GHz, which we use to derive a spectral index of $\alpha = -0.548 \pm 0.024$ for the 2018 Aug 27 observation.

The flux density of J1517-2422 as reported by ALMA was $2.09 \pm 0.06$\,Jy at 343.5\,GHz on 2018 Sep 18, and $2.84 \pm 0.06$\,Jy at 103.5\,GHz on 2018 Sep 20, which we use to derive a spectral index of $\alpha = -0.256 \pm 0.030$ for the 2018 Sep 20 observation.

Because NGC 5643 was observed on two separate days, we combine the MSs using CASA's \texttt{concat} task after calibration and prior to imaging.

\subsection{NGC 5765b}

The targeted flux density calibrator for NGC 5765b, J1550+0527, was quite weak (flux density less than $\sim$0.5\,Jy) at the time of the observation.  We thus use the gain calibrator, J1504+1029, to also calibrate bandpass and flux density for NGC 5765b.  The flux density of J1504+1029 as reported by ALMA on 2018 Nov 11 was $1.96 \pm 0.1$\,Jy at 103.5\,GHz and $0.92 \pm 0.05$\,Jy at 343.5\,GHz, which we use to derive a spectral index of $\alpha = -0.626 \pm 0.060$.

\subsection{CGCG 165-035}

The flux density calibrator for CGCG 165-035 is J1550+0527.  Its flux density as reported by ALMA on 2018 Dec 06 was $0.31 \pm 0.03$\,Jy at 343.5\,GHz, and on 2018 Dec 08 it was $0.90 \pm 0.03$\,Jy at 91.5\,GHz.  We use these values to derive a spectral index of $\alpha = -0.809 \pm 0.078$.

\subsection{NGC 6264}

The flux density calibrator for NGC 6264 is J1751+0939.  Its flux density as reported by ALMA on 2018 Nov 18 was $3.57 \pm 0.06$\,Jy at 103.5\,GHz and $2.08 \pm 0.08$\,Jy at 343.5\,GHz, which we use to derive a spectral index of $\alpha = -0.451 \pm 0.035$.

\section{Spectral line and continuum images at 183\,GHz} \label{app:Images}

\begin{figure*}
    \centering
    \includegraphics[width=0.90\textwidth]{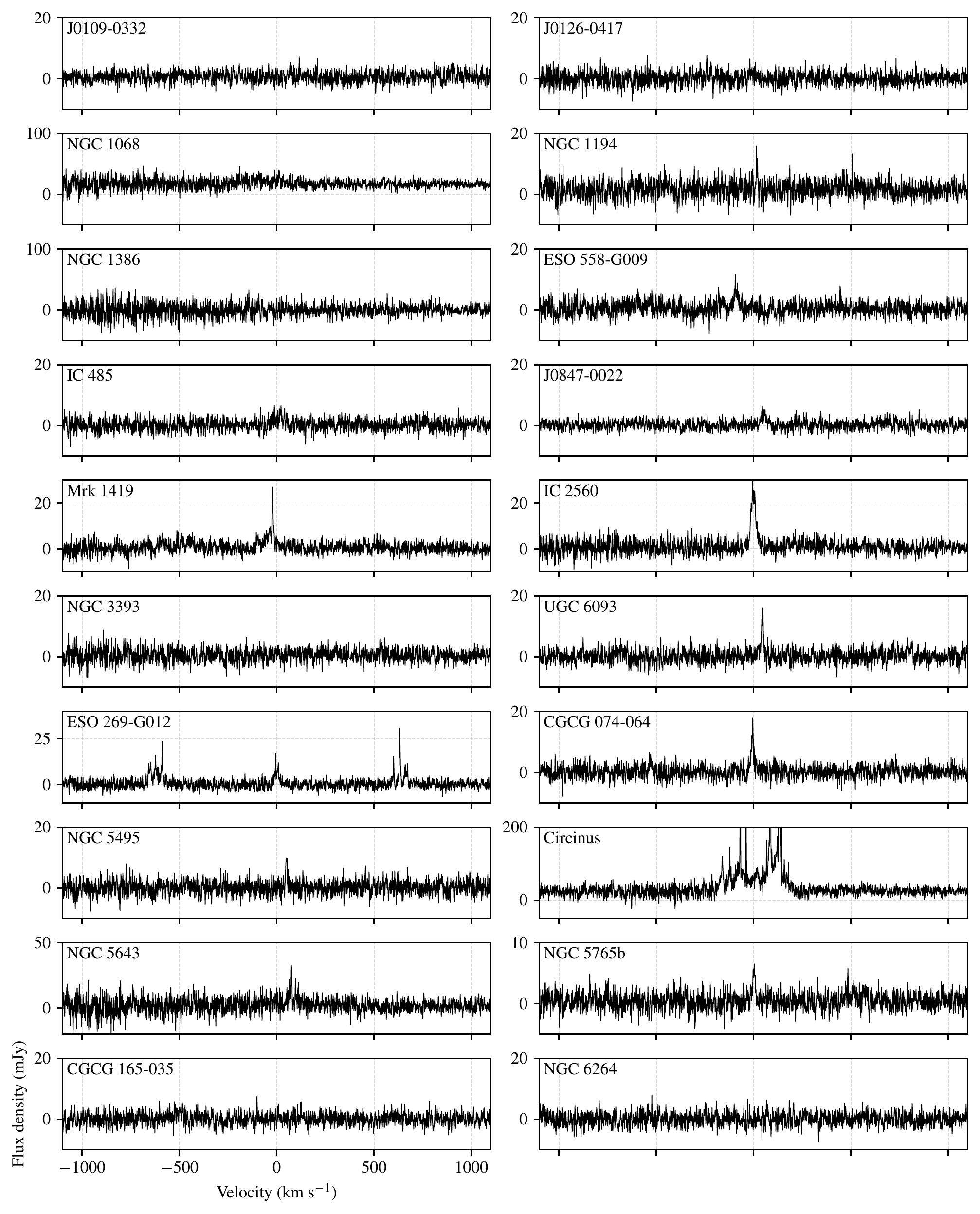}
    \caption{183\,GHz spectra for each of the survey targets, after carrying out boxcar smoothing in frequency by 10 channels (corresponding to a $\sim$2\,\kms post-averaging spectral resolution).  The velocity axis for spectrum is referenced to the recession velocities listed in \autoref{tab:Observations}.  All panels share the same velocity range, which is explicitly labeled in the lower left panel.  We note that the vertical axis range for the Circinus spectrum has been restricted (the strongest emission lines are nearly 5\,Jy) to more clearly show the broad pedestal of emission around the recession velocity.}\label{fig:spectra}
\end{figure*}

\begin{figure*}
    \centering
    \includegraphics[width=0.90\textwidth]{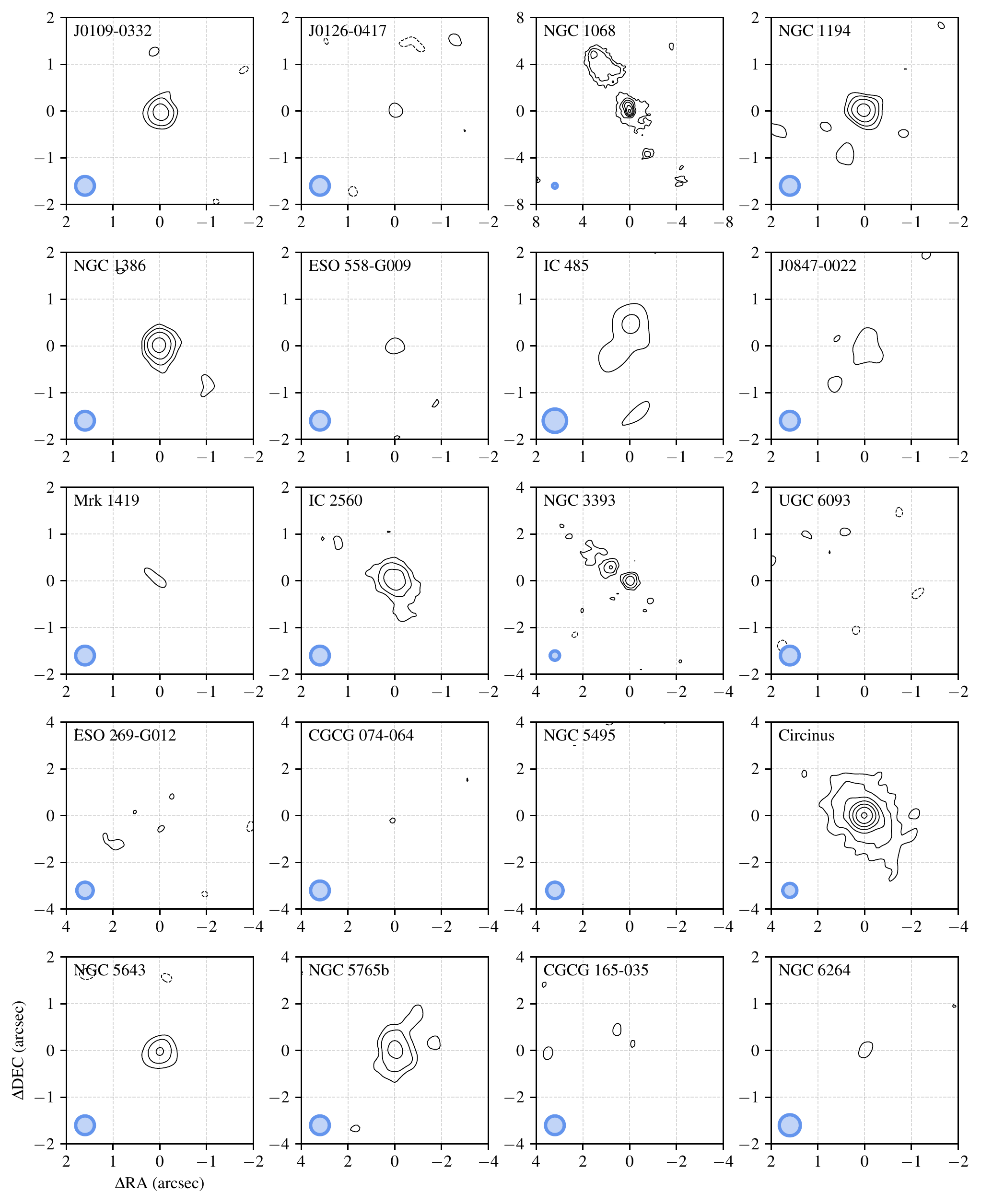}
    \caption{183\,GHz continuum images for each of the survey targets.  The outermost contours mark a brightness level of three times the image RMS (see \autoref{tab:Detections}), and each subsequent contour marks a brightness level that is larger by a factor of two than that of the previous contour.  In each panel, the image is centered on the location of the maser emission (if it is present) or on the peak of the continuum emission (if maser emission is not present); for the one galaxy without either continuum or maser emission detections (CGCG 165-035), we center the image on the expected location of the galaxy (see \autoref{tab:Observations}).  The FWHM restoring beam size is shown in the lower lefthand corner of each panel.}\label{fig:continua}
\end{figure*}

The 183\,GHz spectra for each of our targets are shown in \autoref{fig:spectra}, including those for which no 183\,GHz spectral line emission is detected.  \autoref{fig:continua} shows the corresponding continuum images, again with nondetections included.

\section{Other molecular species in Circinus and NGC 1068} \label{app:HCNetc}

In the continuum spectral windows of both Circinus and NGC 1068, we detect spectral lines other than the 183\,GHz H$_2$O maser line.  We use CASA's \texttt{tclean} task to make image cubes from the continuum spectral windows for both galaxies, and we use the \texttt{curve\_fit} function from the \texttt{scipy} Python library to to fit a Gaussian function to each pixel in the image cube.  We fit for four free parameters at every pixel location, with the parameterization given by

\begin{equation}
S(v) = S_0 + A \exp\left( - \frac{\left( v - v_0 \right)^2}{2 \sigma^2} \right) . \label{eqn:LineGaussian}
\end{equation}

\noindent Here, $S_0$ is a constant-valued background continuum brightness level, $A$ is the peak brightness level of the Gaussian, $v_0$ is the center velocity of the Gaussian, and $\sigma$ is its standard deviation.  We use a least squares optimization scheme to carry out the fits, taking the uncertainty in each pixel's value to be given by the continuum RMS level (see \autoref{tab:Detections}).  The corresponding uncertainty estimates for each modeled parameter are then given by the diagonal elements of the inverse Hessian matrix.

\subsection{Circinus}

\begin{figure*}
    \centering
    \includegraphics[width=1.00\textwidth]{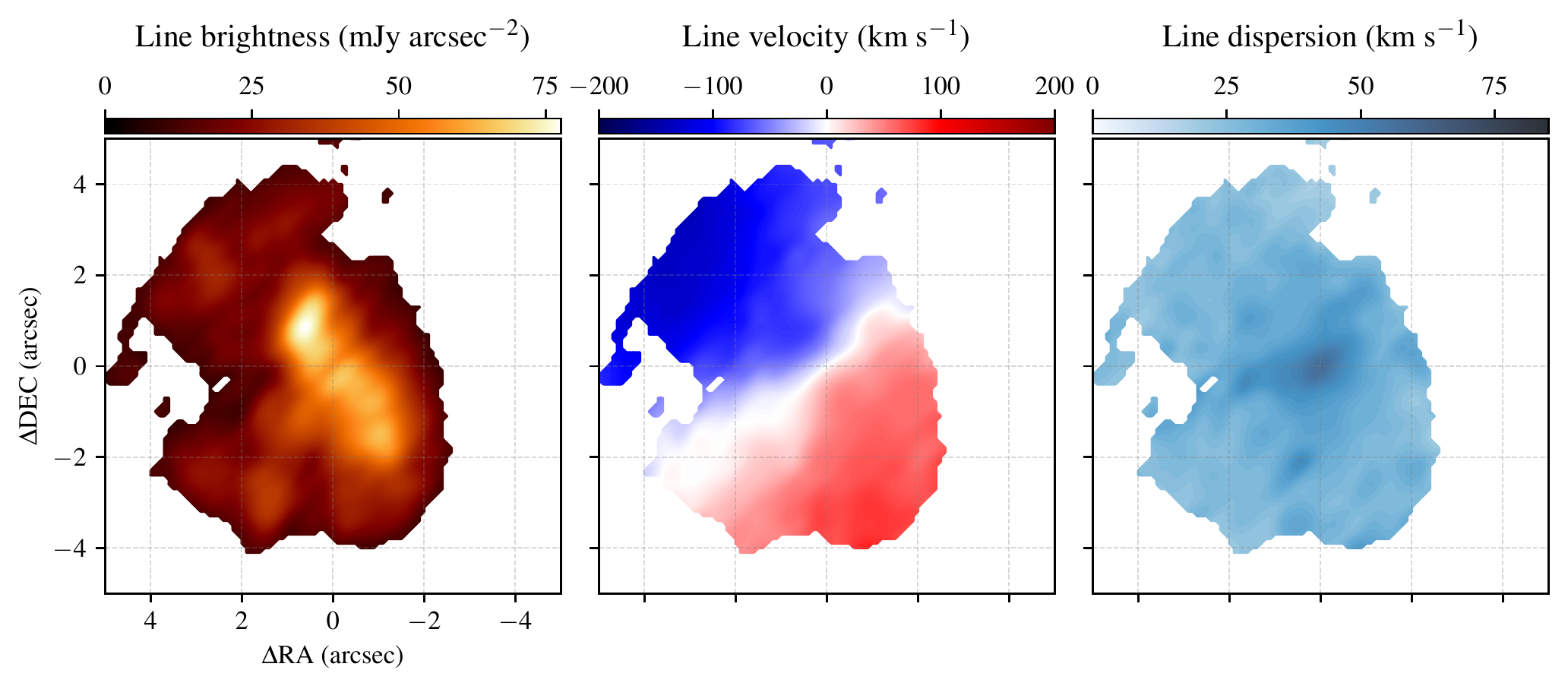}
    \caption{HNC emission in Circinus.  The left panel shows the spatial distribution of the fitted line brightness, the middle panel shows the spatial distribution of the fitted line velocity, and the right panel shows the spatial distribution of the fitted line dispersion; these quantities correspond to the $A$, $v_0$, and $\sigma$ parameters from \autoref{eqn:LineGaussian}, respectively.  Each panel's image is centered as in \autoref{fig:continua}.  We mask out all regions of the image that have a fractional uncertainty in $A$ larger than 20\% ($5\sigma$).}\label{fig:CircinusHNC}
\end{figure*}

There is one additional spectral line detected towards Circinus, which we associate with the HNC ($\text{v}=0$, $\text{J}=2\text{ - }1$) transition with a rest-frame frequency of 181.324758\,GHz \citep{Muller_2005}.  \autoref{fig:CircinusHNC} shows the spatial distributions of the $A$, $v_0$, and $\sigma$ parameters associated with the Gaussian fits to the image cube, which effectively provide an image of the line brightness and its velocity structure.  The emission shows clear and spatially-resolved rotation signatures, with blueshifted material residing in the northeastern region of the galaxy and the redshifted material residing in the southwestern region.  This sense and orientation of rotation matches that seen in large-scale neutral hydrogen maps of Circinus \citep{Jones_1999}, and it also matches that seen in the 183\,GHz maser system presented in this paper (see \autoref{sec:Circinus}).  Other transitions of HNC have previously been observed towards Circinus by \cite{Israel_1992} and \cite{Curran_2001}.

\subsection{NGC 1068}

\begin{figure*}
    \centering
    \includegraphics[width=1.00\textwidth]{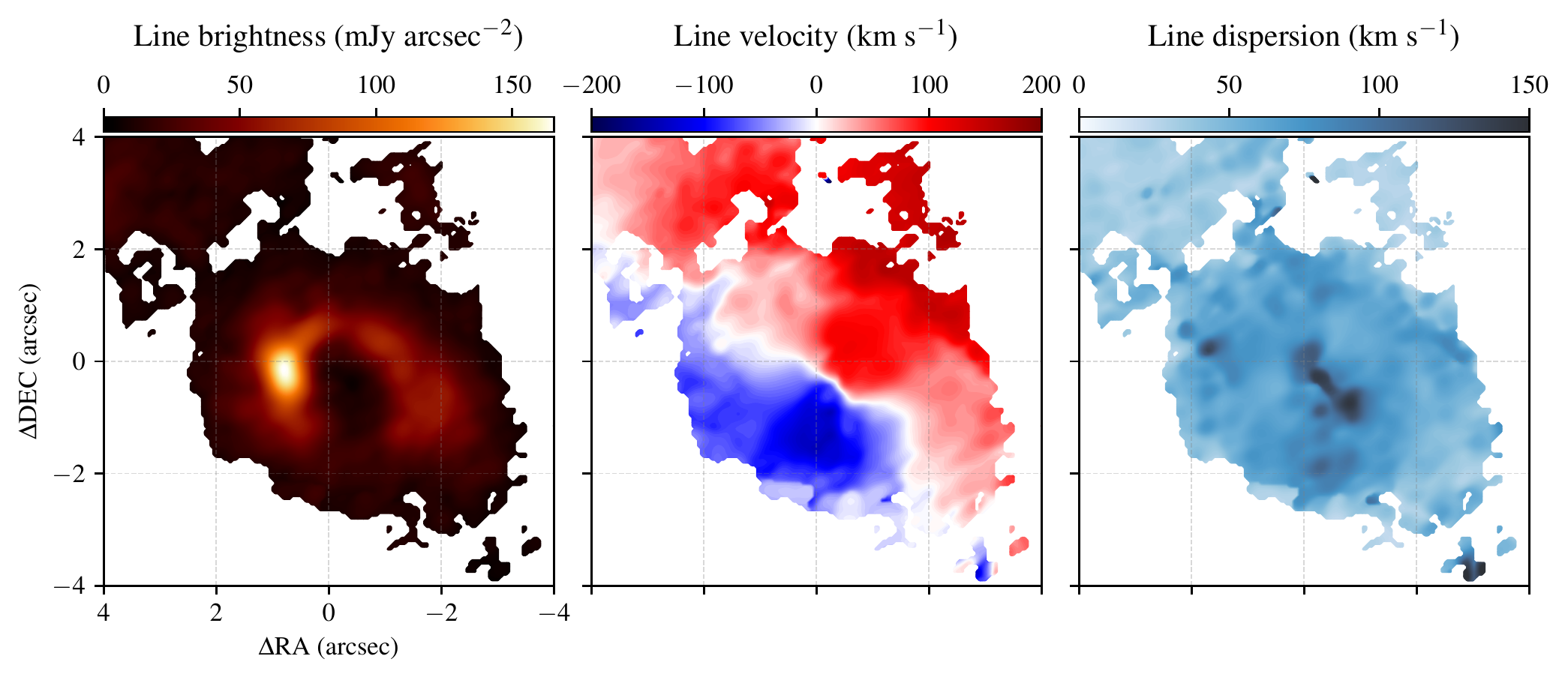}
    \caption{Same as \autoref{fig:CircinusHNC}, but showing HNC emission in NGC 1068.}\label{fig:NGC1068HNC}
\end{figure*}

\begin{figure*}
    \centering
    \includegraphics[width=1.00\textwidth]{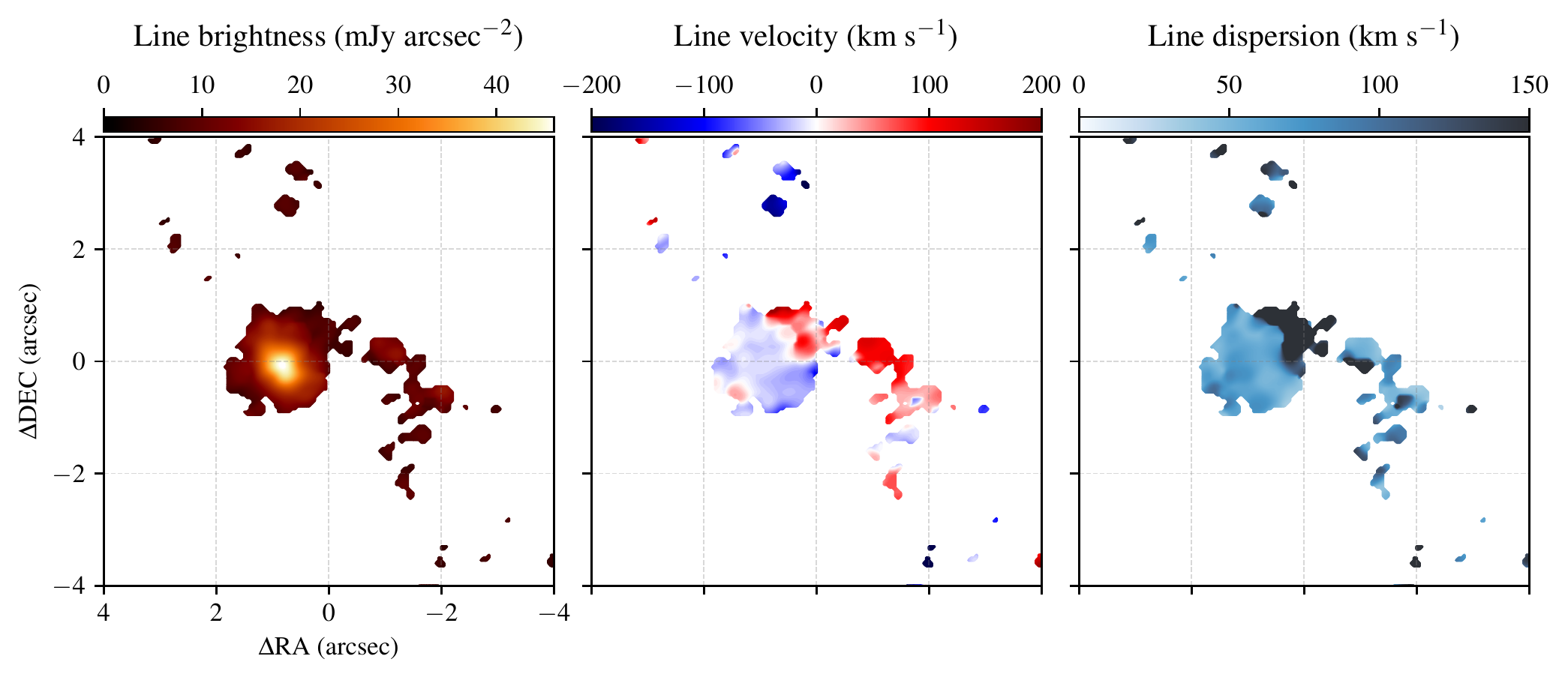}
    \caption{Same as \autoref{fig:CircinusHNC} and \autoref{fig:NGC1068HNC}, but showing HC$_3$N emission in NGC 1068.}\label{fig:NGC1068H3CN}
\end{figure*}

We detect two different lines in the continuum window of the NGC 1068 spectrum, both of which appear to originate in the circumnuclear disk \citep{Schinnerer_2000}.  We associated one of the lines with the same HNC ($\text{v}=0$, $\text{J}=2\text{ - }1$) transition seen towards Circinus; the distribution of the HNC emission is shown in \autoref{fig:NGC1068HNC}.  The other line is weaker and has a more ambiguous identification, but we associate it with the HC$_3$N ($\text{v}=0$, $\text{J}=20\text{ - }19$) transition with a rest-frame frequency of 181.944923\,GHz \citep{Muller_2005}; the distribution of the HC$_3$N emission is shown in \autoref{fig:NGC1068H3CN}.  Both HNC and HC$_3$N have been previously detected towards NGC 1068 \citep[e.g.,][]{Viti_2014,Nakajima_2018,Qiu_2020}.

The emission from both spectral lines shows spatially-resolved rotation signatures, with blueshifted material residing in the southeastern region of the circumnuclear disk and the redshifted material residing in the northwestern region.  The emission is strongest in an eastern ``knot'' within the circimnuclear disk, as seen in previous observations \citep[e.g.][]{Viti_2014,Garcia-Burillo_2019,Qiu_2020}.

\section{Maser position modeling} \label{app:PositionModeling}

We assume that each maser spot is a point source, for which we fit the position $(x,y)$ and flux density $S$ in every spectral channel.  Our model for the point source visibility $V(u,v)$ is given by
\begin{equation}
V(u,v) = S e^{2 \pi i (u x + v y)} .
\end{equation}
We fit directly to the Stokes I complex visibilities in each spectral channel, after self-calibration and averaging in time across the entire observing track.  Because we have self-calibrated using the brightest spectral line at 561.6\,\kms (see \autoref{app:Circinus}), all positions are referenced to the location of the associated maser spot.

Within a single spectral channel, our likelihood function for each data point $i$ is a complex Gaussian \citep{TMS},
\begin{equation}
\ell_i = \frac{1}{2 \pi \sigma^2} \exp\left( - \frac{1}{2 \sigma^2} \left| V_i - \hat{V}_i \right|^2 \right) ,
\end{equation}
where $\hat{V}_i$ represents an observed complex visibility and $\sigma$ is the uncertainty (which we assume to be the same for all data points within a single channel).  Because we do not have a good measure of $\sigma$ from CASA, we treat it as a model parameter and fit for it alongside the position and flux density in each spectral channel.  The total likelihood for a single channel is then the product over all individual data point likelihoods,
\begin{equation}
\mathcal{L} = \prod_i \ell_i .
\end{equation}
We specify uniform priors of $p_S \sim \mathcal{U}(0,10)$ (in units of Jy) for $S$, $p_x \sim \mathcal{U}(-0.5,0.5)$ (in units of arcseconds) for $x$ and $y$, and $p_{\sigma} \sim \mathcal{U}(0,5)$ (in units of Jy) for $\sigma$.
We use the nested sampling package \texttt{dynesty} \citep{Speagle_2020} to carry out the fitting separately for each spectral channel.

\end{document}